\documentclass[a4paper,12pt]{article}
\usepackage[top=1.25in, bottom=1.25in, left=1.05in, right=1.05in]{geometry}
\usepackage[utf8]{inputenc}
\usepackage[T1]{fontenc}
\usepackage[english]{babel}

\usepackage{natbib,longtable}
\usepackage[flushleft]{threeparttable}
\usepackage{threeparttablex} % for "ThreePartTable" environment

\usepackage[section]{placeins}

\usepackage{setspace}

\usepackage{enumitem}
\setlist{noitemsep}  % Reduce space between list items (itemize, enumerate, etc.)

\usepackage[colorinlistoftodos,color=orange!60]{todonotes}
\usepackage{regexpatch}
\makeatletter
\xpatchcmd{\@todo}{\setkeys{todonotes}{#1}}{\setkeys{todonotes}{inline,#1}}{}{}
\makeatother

\usepackage{amscd,amsmath,amssymb,amsfonts,amsthm,bbm,bm,latexsym,mathrsfs,mathtools,bigints}
\usepackage[subnum]{cases}	% multi-case equations with separate number for each
\usepackage{dsfont}
\makeatletter
\newcommand*{\distas}[1]{\mathbin{\overset{#1}{\kern\z@\sim}}}	
\makeatother
\newcommand*\abs[1]{\left|#1\right|}		% absolute value
\allowdisplaybreaks

\theoremstyle{remark}

\theoremstyle{plain}

\RequirePackage[colorlinks=true, citecolor=blue, urlcolor=blue]{hyperref}

\usepackage{xcolor,subfigure,epsfig,epstopdf,rotating,graphics,graphicx} 
\graphicspath{{./figures/}}
\usepackage[width=0.9\textwidth, font={footnotesize}]{caption}

\usepackage{rotating,lscape,pdflscape}
\usepackage{booktabs,longtable,float,array,multirow,colortbl,adjustbox}

\newcolumntype{C}[1]{>{\centering\arraybackslash}p{#1}}

\makeatother

\makeatletter
\def \by{\mathbf{Y}}
\def \by{\mathbf{y}}

\def \bx{\mathbf{x}}
\def \ba{\mathbf{a}}
\def \bu{\mathbf{u}}
\def \bz{\mathbf{z}}
\def \btheta{\boldsymbol{\theta}}

\def \bh{\mathbf{h}}
\def \bbeta{\boldsymbol{\beta}}
\def \bsigma{\boldsymbol{\sigma}}

\def \bmu{\boldsymbol{\mu}}
\def \I {\mathbb{I}}
\def \R {\mathds{R}}

\newcommand{\twostar}{$^{\ast \ast}$}
\newcommand{\threestar}{$^{\ast \ast \ast}$}
\newcommand{\onestar}{$^{\ast}$}

\makeatother

\title{\vspace{-60pt} \textbf{Bayesian Multivariate Quantile Regression with alternative Time-varying Volatility Specifications}
\thanks{\footnotesize
This research used the Computational resources provided by the Core Facility INDACO, which is a project of High Performance Computing at the Università degli Studi di Milano.
}
}

\author{
Matteo Iacopini\thanks{Queen Mary University of London, United Kingdom. \color{blue}\texttt{m.iacopini@qmul.ac.uk}}
\and
Francesco Ravazzolo\thanks{BI Norwegian Business School, Norway and Free-University of Bozen-Bolzano, Italy. \color{blue}\texttt{francesco.ravazzolo@bi.no}}
\and 
Luca Rossini\thanks{University of Milan, Italy. \color{blue}\texttt{luca.rossini@unimi.it}}
}

\date{\today}

\begin{document}

\maketitle

\begin{abstract}
This article proposes a novel Bayesian multivariate quantile regression to forecast the tail behavior of energy commodities, where the homoskedasticity assumption is relaxed to allow for time-varying volatility. In particular, we exploit the mixture representation of the multivariate asymmetric Laplace likelihood and the Cholesky-type decomposition of the scale matrix to introduce stochastic volatility and GARCH processes and then provide an efficient MCMC to estimate them.  
The proposed models outperform the homoskedastic benchmark mainly when predicting the distribution's tails.
We provide a model combination using a quantile score-based weighting scheme, which leads to improved performances, notably when no single model uniformly outperforms the other across quantiles, time, or variables.


\end{abstract}

\clearpage

\onehalfspacing
%\doublespacing

\section{Introduction}      \label{sec:introduction}

The current global economic and financial situation caused by the COVID-19 pandemic and the Russian invasion of Ukraine has renewed the interest of economic forecasters and policy institutions in tail risk. In particular, there has been an increased interest in understanding, modeling, and forecasting the downside tail risk and quantifying the uncertainty around these predictions.
Traditional regression approaches for multivariate economic time series data, such as the vector autoregressive (VAR), model the conditional mean of variables of interest and include time-varying parameters to account for fat tails, deviations from stationarity, and heteroscedasticity.
However, it is often the case that the entire distribution of an indicator (or its quantiles) is relevant, for example, to the policymaker interested in keeping the inflation or the unemployment rates within a specific range. In such cases, focusing exclusively on the conditional mean results in misleading insights as the impact of covariates on the variable of interest may significantly vary across the range of the latter.

To address the limitations of (conditional) mean regression models in capturing the complete picture of the conditional distribution of a multivariate response variable, we propose two novel quantile vector autoregressive (QVAR) models with time-varying volatility.
Specifically, we leverage the quantile regression method \citep[QR, see][]{koenker1978regression} to design a robust framework for the conditional quantiles where the covariates may have a heterogeneous impact across quantile levels.
Conversely, this is a major limitation of mean models, since they consider only the covariates' impact on the conditional mean, which reduces the ability to properly identify the key drivers of the more extreme events (i.e., the tails of the distribution).
%The main advantage of quantile regression over mean models is its ability to investigate the impact of covariates on the entire conditional distribution of the response, instead of considering only the effects on the mean.
%
Besides, we introduce time variation in the scales to capture volatility's persistence and temporal clustering that characterizes economic time series.
Combining these tools allows us to design a coherent framework for comprehensively investigating the entire conditional distribution of multivariate time series.
The proposed quantile QVAR models with a time-varying scale, adopting a quantile regression approach, investigate the conditional quantiles of the response variables, providing a direct answer to the caveats of conditional mean VARs. Moreover, treating extreme observations as related to the tail behavior of the response, the presence of outliers does not contaminate the estimation of all the other quantiles.

%%%%%%%% Quantile AR model (see Carriero et al) + nostro paper %%%%%%%%
With the growing interest in forecasting the entire distribution of economic or energy variables, most of the literature has answered this challenge by relying on a two-block strategy \citep[e.g., see][]{adrian2019vulnerable,carriero2022nowcasting}. They start by specifying a flexible model for the conditional mean, then interpret the quantiles of the implied predictive distribution as the desired quantile forecasts.
In a follow-up, \cite{adrian2020GaR} define the growth-at-risk (GaR) as the effect on the $5$th percentile of conditional growth and estimate it by applying a panel quantile regression model to a collection of eleven advanced economies.
This approach has been extended to density nowcasting, which leverages the informational content of high-frequency data to forecast low-frequency variables in real time. In particular, \cite{ferrara2022Midas} introduced mixed-data sampling (MIDAS) within a Bayesian QR model to make density nowcasts of the GDP. Similarly, \cite{antolin2023advances} proposed a dynamic factor model with stochastic volatility and outlier detection for density nowcasting the growth rates of macroeconomic variables.

The models proposed in this article differ from the previous approaches by adopting a multivariate quantile regression framework replacing a conditional mean model. This choice provides a coherent framework for quantile modeling and forecasting as it relies on the predictive distribution of conditional quantiles, where the latter are directly modeled, allowing heterogeneous covariates' effects. Conversely, the existing methods postulate homogeneity of the covariates' impact by modeling only the conditional mean and using the quantiles of the resulting predictive distribution as an approximation to the predictive of the conditional quantiles.
Moreover, different from scalar-valued methods, the QVAR models with a time-varying scale can capture the contemporaneous cross-sectional dependence characterizing economic and financial time series.

More recent contributions have extended quantile regression to the multivariate setting, originating the quantile vector autoregressive models, among others. \cite{Manganelli2021quantileIRF} introduce a structural QVAR model to capture nonlinear relationships among macroeconomic variables and define a quantile impulse response function to perform stress tests.\footnote{\cite{AB2020}, \cite{CDG2021} and \cite{clark2021investigating} estimate quantile factor models and \cite{korobilis2022probabilistic} extends to probabilistic quantile factor analysis.}
\cite{aastveit2022quantile} develop a forecasting combination scheme from Bayesian quantile regressions that assigns weights to the individual predictive density forecasts based on the quantile scores.
%\cite{iacopini2022bayesian} introduce a novel mixed-frequency quantile VAR model (MF-QVAR), which combines different frequencies in macroeconomic and financial variables to nowcast conditional quantiles of the US GDP.

%%%%%%%% Motivation for time-varying volatility %%%%%%%%
As widely shown in the macroeconomic literature \citep{dAgostino2013SV,clark2015comparison,carriero2016SV}, the distribution of most economic and financial variables typically exhibits significant deviations from Gaussianity. Fat tails, skewness, and time-varying volatility are the most prominent features of these time series data.
To account for them, the standard approach in the literature on conditional mean models relies on the specification of a dynamic process for the volatility of the innovations, such as GARCH or stochastic volatility.
Besides introducing dynamics in the volatility, these approaches can capture fat tails of the marginal distribution of the variable of interest, which helps improve the forecasting performance of economic and financial time series.

Modeling the features mentioned above of the data is crucial, and, in conditional mean models, it has led to fundamental improvements in the forecasting performance of economic, financial, and energy time series \citep{clark2015comparison,gianfreda2023large}.
Despite this success and the renewed interest in quantile forecasting over the last few years, the existing literature on quantile regression has not yet departed from imposing the constraint of constant variance.
This article provides a possible way out of this dilemma by proposing two novel approaches to account for time-varying volatility in a multivariate quantile regression framework.
Besides, a model combination approach is proposed to handle model uncertainty and potential misspecification, leveraging the quantile score to define the models' performance.

%%%%%%%% CONTRIBUTION 1 -- introduce SV + GARCH %%%%%%%%
We aim to fill this gap and propose two frameworks for time-varying volatility in multivariate quantile regression models utilizing parameter-driven and observation-driven specifications. Specifically, we define the likelihood of a QVAR model via the multivariate asymmetric Laplace (MAL) distribution as in \cite{iacopini2022bayesian} and \cite{petrella2019joint}, then we model the conditional variance with either stochastic volatility \citep[SV,][]{cogley2005drifts,Primiceri2005} or generalized autoregressive conditional heteroskedasticity \citep[GARCH,][]{engle1982ARCH, bollerslev1986GARCH}.

%%%%%%%% Computational challenge %%%%%%%%
The asymmetric Laplace distribution \citep{kotz2001laplace} is chosen to represent the likelihood function in quantile regression models; moreover, its representation as a location-scale mixture of Gaussian distributions has allowed the design of efficient estimation algorithms \citep[e.g., see][]{kozumi2011gibbs}.
%But it also complicates the introduction of temporal variation for the scale parameter.
%
However, both proposed extensions have nontrivial computational implications, as coupling SV or GARCH effects with the mixture of Gaussian representation of the asymmetric Laplace distribution results in the volatility also affecting the conditional mean of the conditionally Gaussian distribution, but differently from the traditional SV- or GARCH-in-mean models. Consequently, conventional methods for sampling the volatility path in SV- or GARCH-in-mean models may be inefficient in this quantile framework. As a further contribution, we address this issue by reformulating the models to make possible the joint sampling of the whole trajectory of the time-varying volatility independently along the cross-sectional dimension.

%As widely shown in the macroeconomic literature \citep{dAgostino2013SV, clark2015comparison, carriero2016SV}, introducing alternative time-varying volatility specifications improves the forecasting performance of conditional mean models, particularly in multivariate frameworks.
%However, most of the existing QR models impose constant volatility for the observables, which is a somewhat restrictive assumption when investigating economic and financial time series data; see \cite{GCC2011}, \cite{CAPORIN2018150} and \cite{CAPORIN2021101347} for exceptions in univariate models.

%{\color{red} Aggiungere paragrafo su simulations?}

%%%%%%%% CONTRIBUTION 2 -- compare QVAR models (also with SV + GARCH) %%%%%%%%
Motivated by the lack of a systematic comparison of the performance of quantile regression methods for multivariate time series, a forecasting exercise for different quantiles for several key energy commodities is performed at multiple horizons. We compare multivariate models with and without time-varying volatility. Finally, model combinations are considered based on quantile score weighting schemes \citep{aastveit2022quantile}.

%%%%%%%% Application %%%%%%%%
Specifically, the benchmarks and the proposed QVAR-SV and QVAR-GARCH models are compared in forecasting energy commodities (such as Brent, coal, gas, and CO$_2$) one day and one week ahead.
The results show that the proposed methods beat the constant volatility QVAR benchmark for all the variables investigated; however, no single specification dominates the other ones over time uniformly, nor across variables or quantiles.
Therefore, we propose a model combination with a quantile score-based weighting scheme to provide the forecaster with a practical method that accommodates the vast changes in the forecasting performance of individual models. The combination weights show significant variation over time, especially when the quantiles corresponding to the tails of the distribution are concerned, and at each point in time, most of the mass is assigned to a single model.

The remainder of this article is organized as follows: Section~\ref{sec:model} presents two novel quantile regression models with time-varying volatility.
Section~\ref{sec:estimation} illustrates the Bayesian approach to inference and the combination weights scheme.
Section~\ref{sec:simulation} assesses the performance of the proposed methods in a simulation study.
Then, Section~\ref{sec:application} provides forecasting evidence based on the quantile score for energy commodities data. Finally, Section~\ref{sec:conclusion} concludes the article.

\section{Quantile regression models with alternative time-varying volatility specification}      \label{sec:model}

Let $\by_t \in \R^n$ be an $n$-dimensional vector of response and $\bx_t \in \R^k$ be a $k$-dimensional vector of common covariates.
We remark that maximum likelihood and Bayesian inference based on the multivariate asymmetric Laplace distribution with the appropriate constraints defined in \cite{petrella2019joint} yields an estimate of the conditional of each variable in the system. Specifically, this approach allows to estimate the conditional $\tau_j$-quantile, $Q_{\tau_j}(y_{j,t} | \bx_t)$, for each variable $j=1,\ldots,n$, while also accounting for the cross-sectional dependence among $\by_t$.
In short, the location parameter of the constrained asymmetric Laplace distribution corresponds to the conditional quantile of interest.
Therefore, following the parametrization in \cite{petrella2019joint}, we define a multivariate quantile regression model as:
\begin{equation}
\by_t = B \bx_t + \boldsymbol\epsilon_t, \qquad \boldsymbol\epsilon_t \sim \text{MAL}_n(\mathbf{0}, D \btheta_1, D \Theta_2 \Psi \Theta_2 D),
\label{eq:QR_model}
\end{equation}
where $B$ is a $(n\times k)$ coefficient matrix,  and $\text{MAL}_n(\bmu,\boldsymbol\xi,\Omega)$ denotes the multivariate asymmetric Laplace distribution \citep{kotz2001laplace}, with location $\bmu\in\R^n$, skew parameter $\boldsymbol\xi \in \R^n$, and positive definite scale matrix $\Omega$.
The parametrization of eq.~\eqref{eq:QR_model} is such that $D = \operatorname{diag}(\delta_1,\ldots,\delta_n)$ with variable-specific scales $\delta_j >0$, $\Psi$ is a correlation matrix, and $\Theta_2 = \operatorname{diag}(\btheta_2)$, with:
\begin{equation}
\theta_{1,j} = \frac{1-2\tau_j}{\tau_j(1-\tau_j)}, \qquad
\theta_{2,j} = \sqrt{\frac{2}{\tau_j(1-\tau_j)}}, \qquad j=1,\ldots,n,
\label{eq:theta12}
\end{equation}
where $\tau_j \in (0,1)$ is the (marginal) quantile of the $j$th series.
Building on the mixture representation of the multivariate asymmetric Laplace distribution and defining $\bbeta = \operatorname{vec}(B) \in \R^{nk}$, $\btheta_1 = (\theta_{1,1},\ldots,\theta_{1,n})$, and $X_t = (I_n \otimes \bx_t)$, one obtains:
\begin{align}
\by_t & = X_t \bbeta + D \btheta_1 w_t + \sqrt{w_t} D \Theta_2 \Psi^{1/2} \mathbf{z}_t,   \qquad \mathbf{z}_t \sim \mathcal{N}_n(\mathbf{0},I_n), \quad w_t \sim \mathcal{E}xp(1).
\label{eq:QR_model_mixture}
\end{align}

The multivariate QR in eq.~\eqref{eq:QR_model_mixture} includes the quantile VAR (QVAR) model as a particular case for $\bx_t = \by_{t-1}$. However, it assumes homoskedastic variance for the conditional distribution of the response variables, $\by_t$, which is highly restrictive when modeling economic and financial time series as they are typically characterized by highly persistent and clustered volatility.
The most popular solutions when modeling the conditional mean are the stochastic volatility \citep{taylor1986modelling} and the GARCH \citep{bollerslev1986GARCH} specifications, which belong to the general classes of parameter-driven and observation-driven models \citep{cox1981statistical}, respectively.
To introduce similar dynamics for the time-varying volatility in the conditional distribution of $\by_t$ in eq.~\eqref{eq:QR_model_mixture}, we first reparametrize the model as follows.

From eq.~\eqref{eq:QR_model}, as $D$ and $\Theta_2$ are diagonal matrices, it follows
\begin{equation}
    \Omega = D \Theta_2 \Psi \Theta_2 D = \Theta_2 D \Psi D \Theta_2 = \Theta_2 \Sigma \Theta_2,
\end{equation}
where $\Sigma = D \Psi D$ is a positive definite matrix with $i$th diagonal element given by $\Sigma_{ii}$ and $D = \operatorname{diag}(\Sigma_{11}^{1/2}, \ldots, \Sigma_{nn}^{1/2})$.
% {\color{red} Let us denote by $\Sigma = D \Psi D$ a positive definite matrix, where $D = \operatorname{diag}(\Sigma_{11}^{1/2}, \ldots, \Sigma_{nn}^{1/2})$.}
We assume the Cholesky-type decomposition:
\begin{equation}
\Sigma_t = A H_t A',
\label{eq:Sigma_t}
\end{equation}
where $H_t$ is a diagonal matrix with positive elements on the diagonal and $A$ is a lower triangular matrix with one on the main diagonal. It follows that the (time-varying) elements on the main diagonal of $\Sigma_t$, meaning that is the series-specific conditional variance of $\by_t$, correspond to elements of $H_t$.
Recalling the definition of $D$, one has $H_t^{1/2} = \operatorname{diag}(\Sigma_{t,11}^{1/2},\ldots,\Sigma_{t,nn}^{1/2}) = D_t$.
Therefore, introducing time-varying volatility in the scale matrix, $\Sigma$, of eq.~\eqref{eq:QR_model_mixture} results in a model that includes the square root of the volatility in the conditional mean equation for $\by_t$:
\begin{align}
\by_t & = X_t \bbeta + H_t^{1/2} \btheta_1 w_t + \sqrt{w_t} \Theta_2 A H_t^{1/2} \mathbf{z}_t, \qquad \mathbf{z}_t \sim \mathcal{N}_n(\mathbf{0},I_n), \quad w_t \sim \mathcal{E}xp(1).
\label{eq:QR-TVvolatility}
\end{align}
This represents a generic multivariate quantile regression model with time-varying volatility. In the following sections, we investigate in detail the representation of the matrix $H_t$ when parameter- or observation-driven specifications are considered.

\subsection{Parameter-driven: Stochastic Volatility}      \label{sec:model_SV}

We start by considering the stochastic volatility specification since it has become appealing in macroeconomic time series \citep{clark2015comparison,marcellino2016}. In fact, when dealing with conditional mean multivariate time series models, including stochastic volatility leads to substantial improvements compared to constant volatility models.
This section proposes to model the log volatility as a stationary autoregressive process of order 1. From eq.~\eqref{eq:Sigma_t}, this is obtained by specifying $H_t$ as:
\begin{equation}
\begin{split}
    H_t & = \operatorname{diag}\big( e^{h_{1,t}}, \ldots, e^{h_{n,t}} \big), \\
    h_{j,t} & = \phi_j h_{j,t-1} + \epsilon_{j,t}^h, \qquad \epsilon_{j,t}^h \sim \mathcal{N}(0,\sigma_{h,j}^2),
\end{split}
\label{eq:SV}
\end{equation}
where $\abs{\phi_j} < 1$ and $h_{j,1} \sim \mathcal{N}(0, \sigma_{h,j}^2/(1-\phi_j^2))$.
The model in eq.~\eqref{eq:Sigma_t}-\eqref{eq:QR-TVvolatility}-\eqref{eq:SV} is called the quantile multivariate regression model with stochastic volatility (QR-SV). By introducing lags of the response variable into the covariate set of eq.~\eqref{eq:QR_model_mixture}, one obtains the QVAR-SV model.

By considering the specification of eq.~\eqref{eq:QR-TVvolatility} in eq.~\eqref{eq:QR_model_mixture}, it follows that introducing stochastic volatility in $\Sigma$ results in a model that includes the square root of the volatility terms, $e^{h_{i,t}/2}$, in the conditional mean equation for $\by_t$:
\begin{align}
\by_t & = X_t \bbeta + w_t \Theta_1 e^{\bh_t/2} + \sqrt{w_t} \Theta_2 A \operatorname{diag}(e^{\bh_t/2}) \mathbf{z}_t,
\label{eq:QR-SV_mixture}
\end{align}
where $\bh_t = (h_{1,t},\ldots,h_{n,t})'$, $e^{\bh_t/2} = (e^{h_{1,t}/2},\ldots,e^{h_{n,t}/2})'$, and $\Theta_1 = \operatorname{diag}(\btheta_1)$. The variables $\mathbf{z}_t$ and $w_t$ are as in eq.~\eqref{eq:QR_model_mixture}.
Conditional on $w_t$, this framework resembles a VAR with stochastic volatility in mean (VAR-SVM) model.
In particular, following the notation in \cite{cross2023large}, we have $g = n$ groups with size $n_i = 1$, for each $i=1,\ldots,n$, and $q=0$ lags of the stochastic volatility included in the mean regression.
The main difference is that the VAR-SVM model includes the vector of log-volatilities, $\bh_t$, whereas we have the vector of square roots of volatilities, $e^{\bh_t/2}$.
The main consequence of this difference is that it prevents the use of the sampling scheme in \cite{cross2023large} for estimating the log-volatility process.

To address this challenge and design a computationally efficient procedure for making inference on the log-volatility processes $\bh_j = (h_{j,1},\ldots,h_{j,T})'$, for each series $j=1,\ldots,n$, we start by rewriting the conditional likelihood in eq.~\eqref{eq:QR-SV_mixture}. Notice that in the remaining of this section, we remove $X_t \bbeta$ for notation simplicity, as this term is not essential for delivering the final result.
By rearranging terms in eq.~\eqref{eq:QR-SV_mixture}, one obtains:
\begin{equation}
\by_t = B_t e^{\bh_t/2} + A_t \mathbf{\bar{z}}_t, \qquad \mathbf{\bar{z}}_t \sim \mathcal{N}_n(\mathbf{0},H_t),
\label{eq:transformQR}
\end{equation}
where $B_t = w_t \Theta_1 \in \R^{n \times n}$ and $A_t = \sqrt{w_t} \Theta_2 A \in \R^{n \times n}$ are transformations of $\Theta_1$ and $\Theta_2$. We can notice that $A_t$ is a lower triangular matrix, where the main diagonal elements are different from one. Introducing the inverse of $A_t$ allows us to rewrite eq.~\eqref{eq:transformQR} as
\begin{align}
A_t^{-1} \by_t & = \widetilde{A}_t  e^{\bh_t/2} + \mathbf{\bar{z}}_t, % = \widetilde{A}_t e^{\bh_t/2} + \mathbf{\bar{z}}_t \\
%A_t^{-1} \by_t & = \sum_{i=1}^n \widetilde{A}_{t,:i} e^{h_{i,t}/2} + \mathbf{\bar{z}}_t \\
% A_t^{-1} \by_t - \sum_{i \neq j} \widetilde{A}_{t,:i} e^{h_{i,t}/2} & = \widetilde{A}_{t,:j} e^{h_{j,t}/2} + \mathbf{\bar{z}}_t,
\label{eq:QR-SV-compute_ytilde}
\end{align}
where $\widetilde{A}_t = A_t^{-1} B_t$. Therefore one obtains
\begin{equation}
\widetilde{\by}_t^j = \widetilde{A}_{t,:j} e^{h_{j,t}/2} + \mathbf{\bar{z}}_t, \qquad \mathbf{\bar{z}}_t \sim \mathcal{N}_n(\mathbf{0},H_t),
\label{eq:QR-SV_ytildej}
\end{equation}
where $\widetilde{A}_{t,:j}$ denotes the $j$-th column of $\widetilde{A}_t$ and
\begin{align}
\widetilde{\by}_t^j & = A_t^{-1} \by_t - \sum_{i \neq j} \widetilde{A}_{t,:i} e^{h_{i,t}/2}.
\label{eq:ytilde_j_SV}
\end{align}
Notice that eq.~\eqref{eq:QR-SV_ytildej} is a (conditionally) linear Gaussian regression model with independent response variables, $\widetilde{\by}_t^j$.
Finally, eq.~\eqref{eq:ytilde_j_SV} is used to define the likelihood for the vector $\bh_j$, for each series $j=1,\ldots,n$. With the prior implied by the autoregressive process (independent across $j$), one obtains the posterior full conditional distribution for the entire path of the log-volatility process $\bh_j$.
As this distribution is not of a known family, we use an adaptive random walk Metropolis-Hastings \citep[aRWMH, see][]{atchade2005adaptive} algorithm to draw samples from it.
This approach allows for computationally more efficient sampling of the log-volatility processes, as it substitutes a standard forward loop over time $t$, with a cycle over the series $j$.
As shown in eq.~\eqref{eq:QR-SV_ytildej}-\eqref{eq:ytilde_j_SV}, the likelihood for $h_{j,t}$ depends on on $h_{i,t}$, $i \neq j$ via $\tilde{\by}^j_t$, which prevents from sampling $\bh_j$ in parallel.\footnote{As an approximation, it is possible to sample $\bh_j$ in parallel conditioning on the value of the other $\bh_i$ at the previous iteration of the MCMC.
}%
Therefore, a loop step of computational complexity $O(T)$ is replaced with one of complexity $O(n)$, whose cost can be further reduced by leveraging parallel computing.\footnote{However, this is not the total cost of sampling the entire path of $\bh$, as the latter requires preliminary algebraic operations of cost $O(n^2)$, such as the computation of each $A_t^{-1} \by_t$.}
Furthermore, owing to the autoregressive structure of the process, the joint distribution (over time) of the vector $\bh_j$ has a fast decaying dependence. Overall, this implies that using a Gaussian proposal distribution with a diagonal covariance matrix, $\kappa_j^{(m)} \cdot S_j$, where $S_j$ is diagonal and $\kappa_j^{(m)}>0$ is the series-$j$ adaptive scale at iteration $m=1,\ldots,M$, yields an asymptotic acceptance rate equal to $0.27$.

\subsection{Observation-driven: GARCH}      \label{sec:model_GARCH}

The second time-varying specification we propose is the GARCH process. Despite the vast popularity of the stochastic volatility specification in the recent macro-econometric literature, GARCH models could be considered an alternative specification of time-varying volatility \citep{clark2015comparison}.
The main difference between SV and GARCH models is that conditioning on the information set available at time $t-1$, the SV process is unpredictable due to the inclusion of a fully random innovation. In contrast, the GARCH process is fully predictable, depending only on the previous time observables and latent volatility.
One way to include observation-driven time-varying volatility in eq.~\eqref{eq:Sigma_t} is by assuming a GARCH(1,1) process for each diagonal element of $H_t$ as:
\begin{equation}
\begin{split}
    H_t & = \operatorname{diag}\big( \sigma_{1,t}^2, \ldots, \sigma_{n,t}^2 \big), \\
    \sigma_{j,t}^2 & = \omega_j + \alpha_j \epsilon_{j,t-1}^2 + \gamma_j \sigma_{j,t-1}^2 \\
                   & = \omega_j + \alpha_j \big( y_{j,t-1} -X_{j,t-1}\bbeta -w_{t-1} \theta_{1,j} \sigma_{j,t-1} \big)^2 + \gamma_j \sigma_{j,t-1}^2,
\end{split}
\label{eq:GARCH}
\end{equation}
where the parameters need to follow the following constraints to ensure stationarity: $\omega_j >0$, $\alpha_j \geq 0$, $\gamma_j \geq 0$, and $(\alpha_j + \gamma_j) < 1$.
The model in eq.~\eqref{eq:QR_model_mixture}-\eqref{eq:Sigma_t}-\eqref{eq:GARCH} is called the quantile multivariate regression with generalized autoregressive conditional heteroskedasticity model (QR-GARCH). As in the previous case, one obtains the quantile multivariate VAR model with GARCH (QVAR-GARCH) by including lags of the response variable among the covariates.

Plugging \eqref{eq:GARCH} into eq.~\eqref{eq:QR-TVvolatility} results as in the SV case in a model including the square root of the volatility in the mean equation for $\by_t$ as
\begin{align}
\by_t & = X_t \bbeta + w_t \Theta_1 \bsigma_t + \sqrt{w_t} \Theta_2 A \operatorname{diag}(\bsigma_t) \mathbf{z}_t,
\label{eq:QR-GARCH_mixture}
\end{align}
where $\bsigma_t = (\sigma_{1,t},\ldots,\sigma_{n,t})'$.
Conditional on $w_t$, this framework resembles a VAR with GARCH in mean (VAR-GARCH-M) model. The main difference is that the VAR-GARCH-M model includes the vector of volatilities, $\bsigma_t^2$, whereas we have the vector of square roots of volatilities, $\bsigma_t$.

As the stochastic volatility scenario states, we remove $X_t \bbeta$ for notational simplicity. Then, following the approach defined in eq.~\eqref{eq:QR-SV-compute_ytilde}, one obtains:
%\begin{equation}
%\by_t = B_t \bsigma_t + A_t \mathbf{\bar{z}}_t \qquad \mathbf{\bar{z}}_t \sim \mathcal{N}_n(\mathbf{0},H_t),
%\end{equation}
%then
%\begin{align}
%A_t^{-1} \by_t & = \widetilde{A}_t \bsigma_t + \mathbf{\bar{z}}_t %\\
%%A_t^{-1} \by_t - \sum_{i \neq j} \widetilde{A}_{t,:i} \sigma_{i,t} & = \widetilde{A}_{t,:j} \sigma_{j,t} + \mathbf{\bar{z}}_t,
%\label{eq:QR-GARCH-compute_ytilde}
%\end{align}
%and finally 
\begin{equation}
\widetilde{\by}_t^j = \widetilde{A}_{t,:j} \sigma_{j,t} + \mathbf{\bar{z}}_t, \qquad \mathbf{\bar{z}}_t \sim \mathcal{N}_n(\mathbf{0},H_t).
\end{equation}
where 
\begin{align}
\widetilde{\by}_t^j & = A_t^{-1} \by_t - \sum_{i \neq j} \widetilde{A}_{t,:i} \sigma_{i,t}.
\label{eq:ytilde_j_GARCH}
\end{align}
Similarly to eq.~\eqref{eq:ytilde_j_SV} for the stochastic volatility case, eq.~\eqref{eq:ytilde_j_GARCH} allows to single out the contribution of the $j$th time-varying variance to the likelihood, thus allowing to design an efficient algorithm to make inference on $\sigma_{j,t}^2$ and the associated static parameters, for each series $j=1,\ldots,n$.

\section{Model Estimation and Evaluation}        \label{sec:estimation}

\subsection{Bayesian Inference}
This section provides the details of our proposed multivariate quantile regression model estimation with stochastic volatility and GARCH error terms. 

We start the description from the common parameters across the QR-SV and QR-GARCH models, that is, the vectorized matrix of coefficients, $\bbeta$, and the vector containing the non-zero elements of the $j$-th row of the $A$ matrix, $\overline{\ba}_j$. We assume the following prior distributions:
\begin{align}
\bbeta  \sim \mathcal{N}_{nk}(\underline{\bmu}_\beta, \underline{\Sigma}_\beta), \qquad
\overline{\ba}_j \sim \mathcal{N}_{j-1}(\underline{\bmu}_{a,j}, \underline{\Sigma}_{a,j}), \qquad j=2,\ldots,n,
\end{align}
where the hyperparameters are chosen such as to be noninformative. This structure can be easily extended in high-dimensional settings to allow for global-local shrinkage priors. However, this is beyond the scope of this article.

For the QR-SV model, the prior specification on the parameters of the log-volatility equation follows the setup proposed in \cite{kastner2014ancillarity}. For each series $j=1,\ldots,n$, we specify the following prior distributions for the (transformed) persistence parameter and the innovation variance of the log-volatility process:
\begin{align}
\Big( \frac{1+\phi_j}{2} \Big) \sim \mathcal{B}e(\underline{a}_\rho, \underline{b}_\rho), \qquad 
\sigma_{h,j}^2 \sim \mathcal{IG}(\underline{a}_\sigma, \underline{b}_\sigma).
\end{align}
For the QR-GARCH model, we assume a Gaussian prior for the log transformation of the parameters and impose the stationarity condition by truncating the support of the joint prior $(\alpha_j,\gamma_j)$ as follows:
\begin{align}
\begin{split}
    \log(\omega_j) & \sim \mathcal{N}(\underline{\mu}_\omega, \underline{\sigma}_\omega^2) \\
    \left( \begin{array}{c}  \log(\alpha_j) \\ \log(\gamma_j) \end{array} \right) & \sim \mathcal{N}_2\left( \left( \begin{array}{c}
    \underline{\mu}_\alpha \\  \underline{\mu}_\gamma
\end{array} \right), \left( \begin{array}{cc}
    \underline{\sigma}_\alpha^2 & 0 \\ 0 & \underline{\sigma}_\gamma^2
\end{array} \right) \right) \I(\alpha_j+\gamma_j < 1).
\end{split}
\end{align}

As the joint posterior distribution is not tractable, we design an efficient Markov chain Monte Carlo (MCMC) algorithm based on the Metropolis within Gibbs sampler to approximate it.
We refer to Appendix~\ref{S_sec:post_condit_QVAR_SV} and \ref{S_sec:post_condit_QVAR_GARCH} for the details of the full conditional distributions of the parameters for all the models. Concerning the trajectory of the time-varying volatility from the QR-SV model, the result in eq.~\eqref{eq:ytilde_j_SV} allows us to design a sampling scheme for the whole path of $\bh_j$, independently for each series $j=1,\ldots,n$.
In particular, the path of $\bh_j$ is drawn using an adaptive random walk Metropolis-Hastings (aRWMH) algorithm with a Gaussian proposal given by:
\begin{equation*}
\bh_j^* \sim \mathcal{N}\big( \bh_j^{(m-1)}, \kappa_j^{(m-1)} S_j \big),
\end{equation*}
where $S_j$ is a fixed covariance matrix and $\kappa_j^{(m-1)}$ is a scalar scale parameter that is adapted over the MCMC iterations $m=1,\ldots,M$ as in \cite{atchade2005adaptive}.
This approach replaces a forward cycle over $t=1,\ldots,T$ with an (independent) cycle over $j=1,\ldots,n$, reducing the computational cost as typically $n < T$. Moreover, as the dependence between the elements of the vector $\bh_j$ rapidly decays, using a diagonal proposal covariance $S_j$ yields satisfactory results.
On the other hand, the static parameters,  $\omega_j$, $\alpha_j$, and $\gamma_j$, of the GARCH process in the QR-GARCH model, for $j=1,\ldots,n$, are sampled using an aRWMH algorithm with a log-normal proposal. Then, conditional on a draw of $\omega_j$, $\alpha_j$, $\gamma_j$, the path of the volatility $\bsigma_j^2 = (\sigma_{j,1}^2,\ldots,\sigma_{j,T}^2)$ is computed.
The asymptotic acceptance rates of the aRWMH steps have been set to $0.30$.

\subsection{Evaluation and combination}

To assess the quality of the quantile forecasts, we rely on the quantile score \citep[QS, see][]{giacomini2005evaluation,carriero2022nowcasting} as a tail metric. The QS for model $k=1,\ldots,K$, where $K$ is the total number of models estimated in the forecasting exercise, at forecasting horizon $h=1,\ldots,H$ and quantile $\tau$, is defined as:
\begin{align*}
    \text{QS}_{k,\tau,t+h} = (\by_{t+h} - \hat{Q}_{k,\tau,t+h}) \odot (\tau - \mathbb{I}_{\{\by_t \leq \hat{Q}_{k,\tau,t+h}\}}),
\end{align*}
where $\odot$ denotes the Hadamard product, $\by_{t+h}$ is the observed value of the vector response to be forecasted, $\hat{Q}_{k,\tau,t+h}$ is the forecast of quantile $\tau$ under model $k$, and $\I_{\{ C \}}$ is a vector-valued indicator function, whose $j$th element has value of $1$ if the outcome $y_{j,t+h}$ is at or below the forecasted quantile $\hat{Q}_{j,k,tau,t+h}$ and $0$ otherwise.
Notice that better performances are associated with lower values of the QS.

Besides model evaluation, we follow \cite{aastveit2022quantile} and propose a combination of different models based on the QS. In particular, we provide two different forecast combinations based on time-varying weights (T-V) and constant (average) weights (AVG). Concerning time-varying weights, we define the weight of model $k$ at forecasting horizon $h$ and quantile $\tau$ as:
\begin{equation}
    w_{k,\tau,t+h} = \frac{\sum_{t=t_i}^{t_i+t_o-h} QS_{k,\tau,t}^{-1}}{\sum_{j=1}^K \sum_{t=t_i}^{t_i+t_o-h} QS_{j,\tau,t}^{-1}},
\label{eq:QS_weigths}
\end{equation}
where $t_i$ and $t_o$ are the in-sample and out-of-sample periods, respectively. Therefore, the weights are a function of the past performance of each model $k$ known when the forecast is made.
The resulting forecast combination is obtained as the time-varying weighted average:
\begin{equation}
Q_{\tau,t+h}^{c,tv} = \sum_{k=1}^K w_{k,\tau,t+h} \times QS_{k,\tau,t+h},
\label{eq:y_forec_combine}
\end{equation}
where the acronym ${tv}$ states for time-varying weights. As a second weighting scheme, we consider the forecast combination obtained by using the temporal average of the weights defined in eq.~\eqref{eq:QS_weigths}, that is:
\begin{equation}
    \bar{w}_{k,\tau} = \frac{1}{t_o}\sum_{t=1}^{t_o} w_{k,\tau,t+h},
\end{equation}
which yields the average forecast combination:
\begin{equation}
    Q_{\tau,t+h}^{c,avg} = \sum_{k=1}^K \bar{w}_{k,\tau} \times QS_{k,\tau,t+h},
\end{equation}
where the acronym $avg$ states for weight average, and in the following tables, we use Q(V)AR Combination (T-V) and (AVG) to define the time-varying and average weights, respectively.

To compare the performances across the different models, we apply the Diebold-Mariano $t$-test \citep{Diebold1995} for equality of the quantile scores to compare the predictions of alternative models with the benchmark for a given horizon $h$. The differences in accuracy that are statistically different from zero are denoted with one, two, or three asterisks, corresponding to significance levels of $10\%$, $5\%$, and $1\%$, respectively. Our use of the Diebold-Mariano test, with forecasters from often nested models, is a deliberated choice as in \cite{clark2015comparison}. We report the $p$-values based on one-sided tests, taking the QVAR as the null and the other current models as alternatives.
Finally, we applied the model confidence set procedure of \cite{hansen_etal.2011} across models for a fixed horizon to jointly compare their predictive power without disentangling multivariate models and their combinations.

\section{Simulation studies}
\label{sec:simulation}

This section is devoted to highlighting the advantages of including time-varying volatility over the constant volatility benchmark in multivariate quantile time series models across different simulation studies. Our main purpose is to illustrate that the proposed Bayesian QVAR models with stochastic volatility or GARCH behave better than Bayesian QVAR models with constant volatility across different quantiles. 

We carry out a Monte Carlo simulation study where we generate the data from a VAR(1) model with stochastic volatility and skew-$t$ innovations. In particular, we generate $T = 200$ observations of $n=9$ variables and we consider for estimation purposes the quantiles $\tau = 0.1, 0.2, 0.3, 0.4, 0.5, 0.6, 0.7, 0.8, 0.9$. For each quantile level, we run $15$ independent replications of the MCMC and focus on the median of the mean absolute deviation (MMAD), that is $\text{median}\Big( T^{-1} \sum_{t=1}^T |\textbf{x}_t' \widehat{\bbeta} - \textbf{x}_t' \bbeta| \Big)$, where $\widehat{\bbeta}$ is the posterior mean of the coefficients. Since this measure refers to the conditional quantile, as a second measure of interest we use the Frobenius norm (FN) of the vectorised coefficient matrix $\bbeta$.

Table~\ref{tab:simu} reports the two measures of interest for the QVAR benchmark model across different quantiles. For the two time-varying volatility models, it provides the ratio of each measure's value over the benchmark, such that a value smaller than one means that the QVAR-SV or QVAR-GARCH is outperforming the QVAR.
%On the other hand, if the ratio is less than 1, it means that including time-varying volatility provides improvements against the constant volatility specification.

\begin{table}[t!h]
    \centering
    \setlength{\tabcolsep}{3.2pt}
\begin{tabular}{l|cc|cc|cc|cc|cc}
    \hline
  Models  & \multicolumn{2}{c|}{$\tau = 0.1$} & \multicolumn{2}{c|}{$\tau = 0.2$} & \multicolumn{2}{c|}{$\tau = 0.3$} & \multicolumn{2}{c|}{$\tau = 0.4$} & \multicolumn{2}{c}{$\tau = 0.5$} \\
    & MMAD & FN & MMAD & FN  & MMAD & FN & MMAD & FN & MMAD & FN \\
    QVAR & 0.349 &  0.023 & 0.345 & 0.022 & 0.340 & 0.021 & 0.338 & 0.021 & 0.340 & 0.022 \\
    QVAR-SV & 0.955 & 0.899 & 0.981 & 0.935 & 0.973 & 0.948 & 0.980 & 0.969 & 0.994 & 0.971\\
    QVAR-GARCH & 1.010 & 1.059 & 0.988 & 0.971 & 0.938 & 0.898 & 0.967 & 0.933 &  0.910 & 0.834\\
    \hline \\[-0.4cm]
  Models  & \multicolumn{2}{c|}{$\tau = 0.6$} & \multicolumn{2}{c|}{$\tau = 0.7$} & \multicolumn{2}{c|}{$\tau = 0.8$} & \multicolumn{2}{c|}{$\tau = 0.9$} & & \\ 
      & MMAD & FN & MMAD & FN  & MMAD & FN & MMAD & FN & & \\
%\\[-0.4cm]
\hline 
    QVAR & 0.345 & 0.022 & 0.346 & 0.023 & 0.349 & 0.023 & 0.351 & 0.024 \\    
    QVAR-SV &  0.976 & 0.959 & 0.956 & 0.910 & 0.969 & 0.931 & 0.951 & 0.909\\
    QVAR-GARCH & 0.945 & 0.937 & 1.028 & 1.023 & 0.992 & 0.981 & 1.072 & 1.177\\
    \hline 
    \end{tabular}
    \caption{Median mean absolute deviation (MMAD) and Frobenius norm (FN) for the QVAR model at different quantile levels $\tau$. The second and third rows present the ratios of the models with respect to the QVAR, where values smaller than 1 mean the proposed model outperforms the QVAR model.}
    \label{tab:simu}
\end{table}

As shown in Table~\ref{tab:simu}, the proposed QVAR models with time-varying volatility perform better in the tails compared to the center of the distribution. In particular, in the left tails ($\tau = 0.1$), the QVAR-SV improves by 5\% and 10\%  based on the two metrics considered. Moving on to the right tails ($\tau = 0.9$), the situation is similar regarding the QVAR-SV with improvements in the order of 5\% and 10\% with respect to the benchmark model. On the other hand, the QVAR-GARCH model seems to have a more unstable performance across the quantiles, indeed in some quantiles ($\tau = 0.3, 0.5$), it improves by 10\% by looking at the FN metric with respect to the constant volatility model, while it seems to have less improvements in the right tails with respect to the stochastic volatility model.

%, while the QVAR-GARCH improvement seems less relevant. These results also hold in the right tails ($\tau =0.8$), where the inclusion of SV improves by 5\% and 10\%, while GARCH gains 8\% and 19\%.

%%%%%%%%%%%%%%%%%%%%%%%%%%%%%%%%%%%%%%%%%%%%%%%%%%%%%%%%%%%%%%%%%%%%%%%%%%%

\section{Application}        \label{sec:application}

We use four daily energy commodities as in \cite{Ravazzolo23}: Brent, $CO_2$, coal, and natural gas prices related to the European markets. All the variables are converted into euros based on the US dollar-to-euro exchange rate. ICE Brent represents the future oil price traded in dollars, while $CO_2$ represents the carbon emissions for the Euro area. Coal is based on the price of coal delivered to the Netherlands and Belgium, and the natural gas price (Dutch TTF) is the hub benchmark for the European market.
The data starts on 01 January 2021 and ends on 30 December 2022,\footnote{As a robustness check, we have considered different years (2020, 2021, and 2023) and we reported the results in the Supplement.} and we model the growth rates of the prices, computed as $100\cdot(x_{j,t}-x_{j,t-1})/x_{j,t-1}$, where $x_{j,t}$ is the price of variable $j$ at time $t$.
Table~\ref{tab:data_2022_summary} reports each series's variance, skewness, and kurtosis, providing evidence of substantial deviations from Gaussinity. Finally, the variables have then been standardized to ease numerical computations and facilitate comparisons of the estimated coefficients.

\begin{table}[t!h]
    \centering
    \begin{tabular}{c|cccc}
    \hline
    Measure & Brent & $CO_2$ & Coal & Gas \\
    \hline
    \multicolumn{2}{c}{\textit{Full sample (2021-2022)}} \\
    Variance & 6.790 & 9.322 & 26.734 & 96.918\\
    Skewness & 0.936 & 0.992 & 5.209 & 0.712 \\
    Kurtosis & 6.313 & 8.869 & 75.919 & 12.616 \\
        \hline
    \multicolumn{2}{c}{\textit{Year 2021}} \\
    Variance & 4.361 & 7.763 & 28.668 & 38.732 \\
    Skewness & 1.141 & 1.239 & 9.158 & 1.097 \\
    Kurtosis & 7.925 & 7.994 & 120.144 & 8.456 \\
        \hline
    \multicolumn{2}{c}{\textit{Year 2022}} \\
    Variance & 9.245 & 10.849 & 24.895 & 155.025 \\
    Skewness & 0.776 & 0.784 & 0.286 & 0.464 \\
    Kurtosis & 5.037 & 9.021 & 16.035 & 9.311\\
    \hline
    \end{tabular}
    \caption{Summary statistics for the period from 01 January 2021 to 30 December 2022.}
    \label{tab:data_2022_summary}
\end{table}

In light of the critical role that time-varying volatility will play in our forecasting results, we start by looking at the movements of the volatility for some specific variables and quantiles. We use a quantile VAR model with one lag (with and without time-varying volatility) and consider 17 quantiles ranging from 0.1 to 0.9 (equally spaced). In the following, for simplicity, we provide the results for some selected quantiles and refer to the Supplement for the results about all the other quantiles.

In the real-time out-of-sample forecast exercise, we assess and compare all the models through the quantile score. To get further insights, we also report the associated model combination weights from eq.~\eqref{eq:QS_weigths} over time. As the data are collected daily, we use a rolling window approach for the forecasting exercise and set the length of the in-sample period to $261$ days. Therefore, the out-of-sample period starts on 03 January 2022 and ends on 23 December 2022.

\begin{figure}[t!h]
\centering
\hspace*{-2.8ex}
\begin{tabular}{c ccc}
 & {\small $\tau = 0.1$} & {\small $\tau = 0.5$} & {\small $\tau = 0.9$} \\
\begin{rotate}{90} \hspace{30pt} {\scriptsize Brent} \end{rotate} \hspace*{-10pt} &
\includegraphics[trim= 0mm 0mm 0mm 0mm,clip, width= 5.0cm]{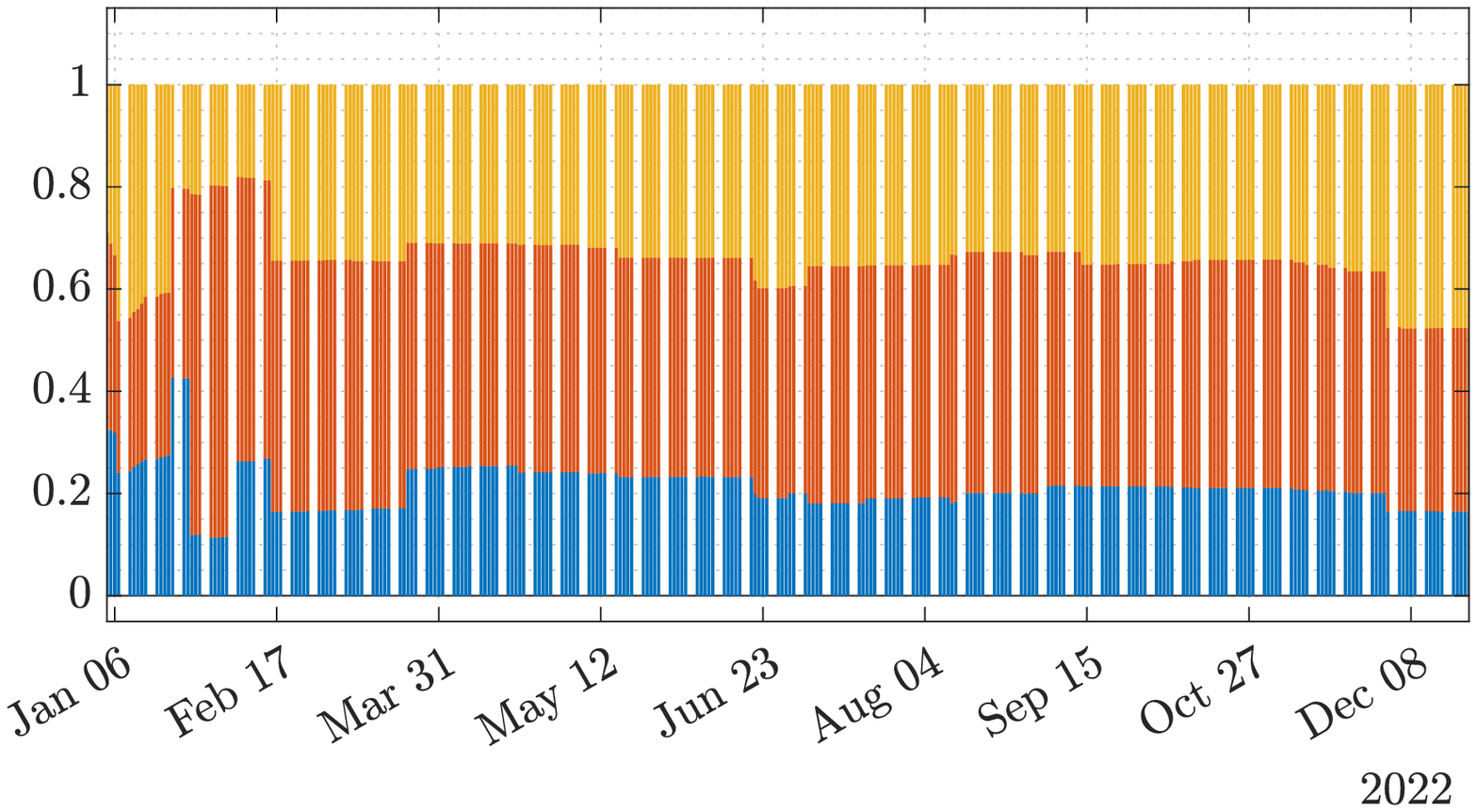} &
\includegraphics[trim= 0mm 0mm 0mm 0mm,clip, width= 5.0cm]{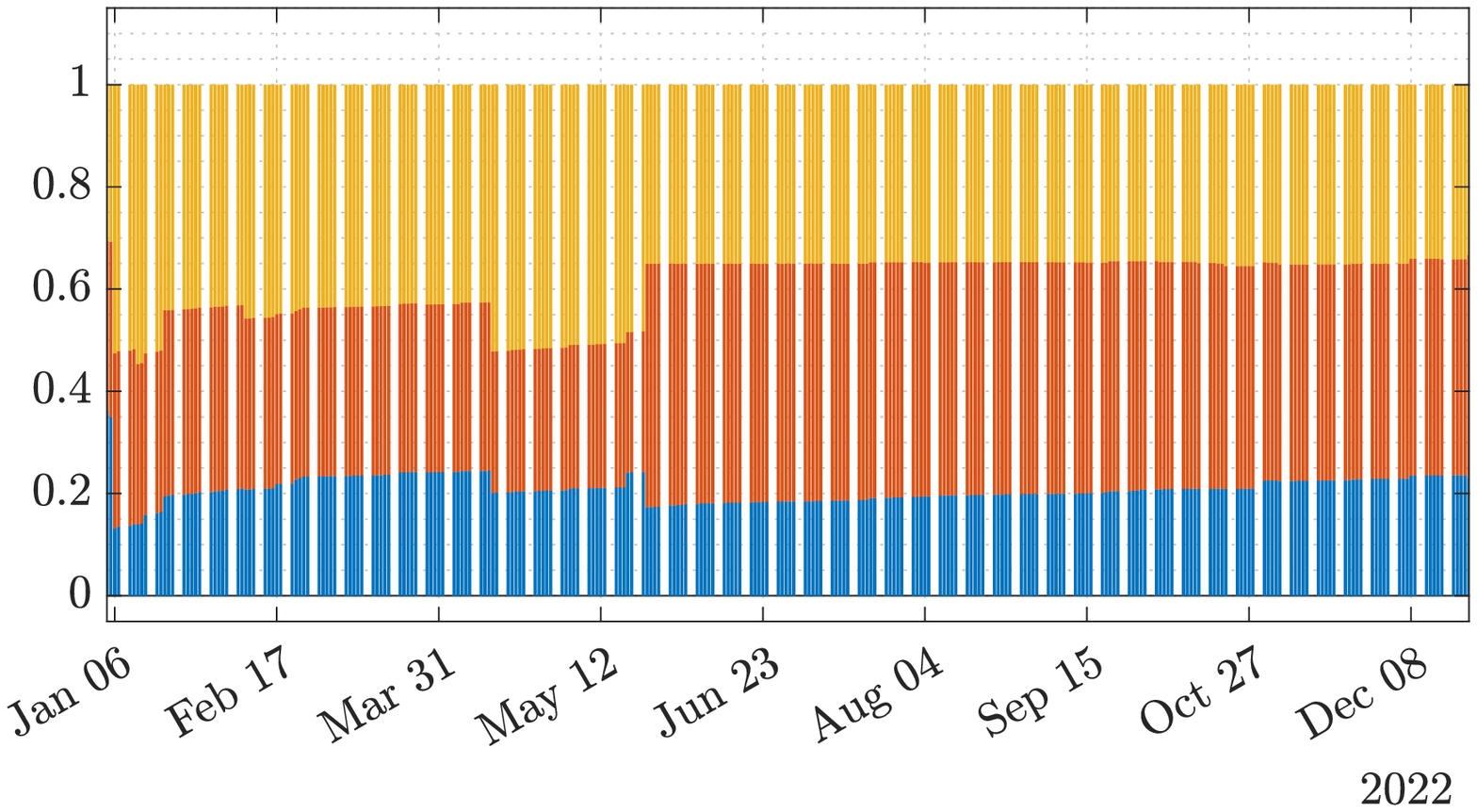} &
\includegraphics[trim= 0mm 0mm 0mm 0mm,clip, width= 5.0cm]{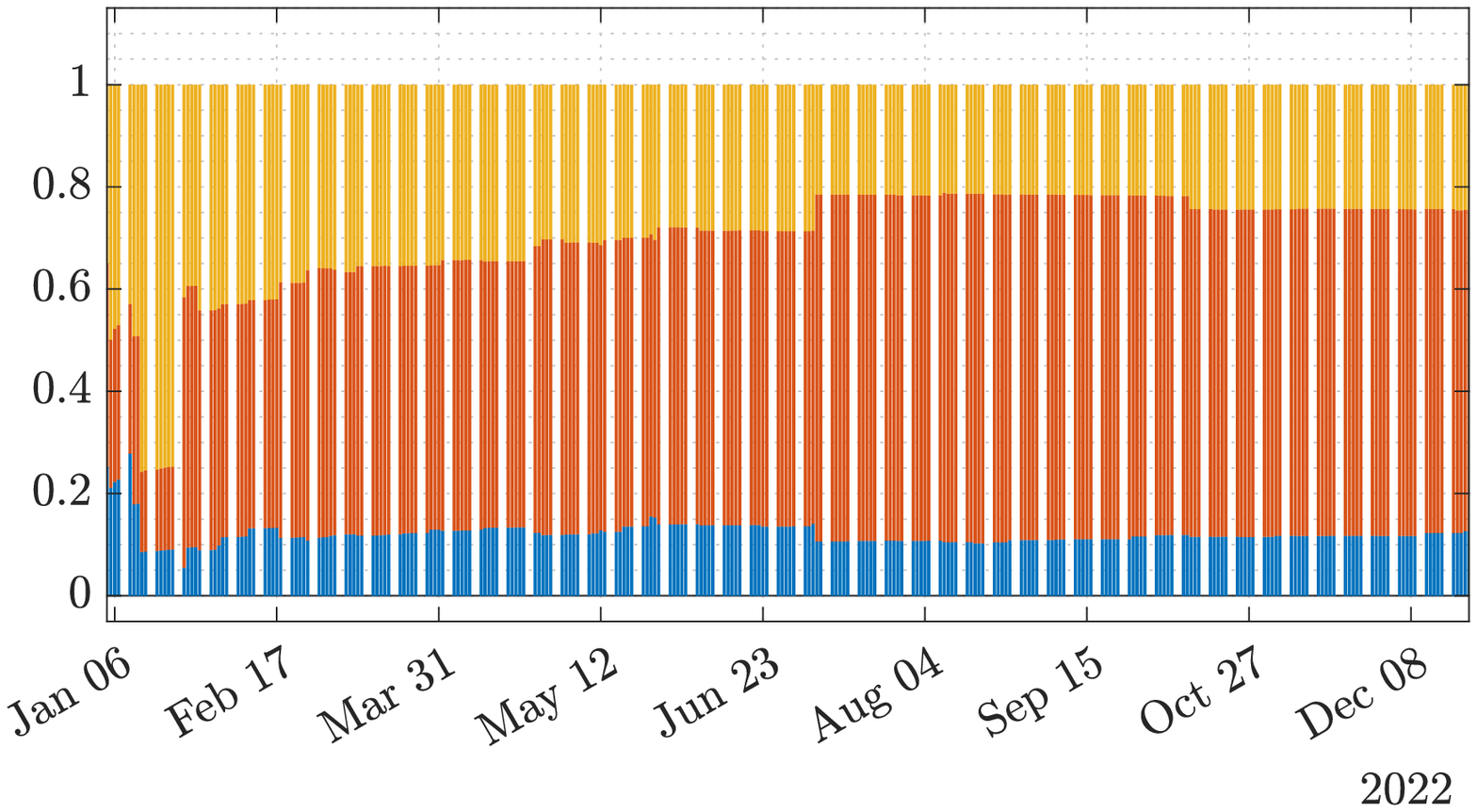} \\
\begin{rotate}{90} \hspace{30pt} {\scriptsize CO2} \end{rotate} \hspace*{-10pt} &
\includegraphics[trim= 0mm 0mm 0mm 0mm,clip, width= 5.0cm]{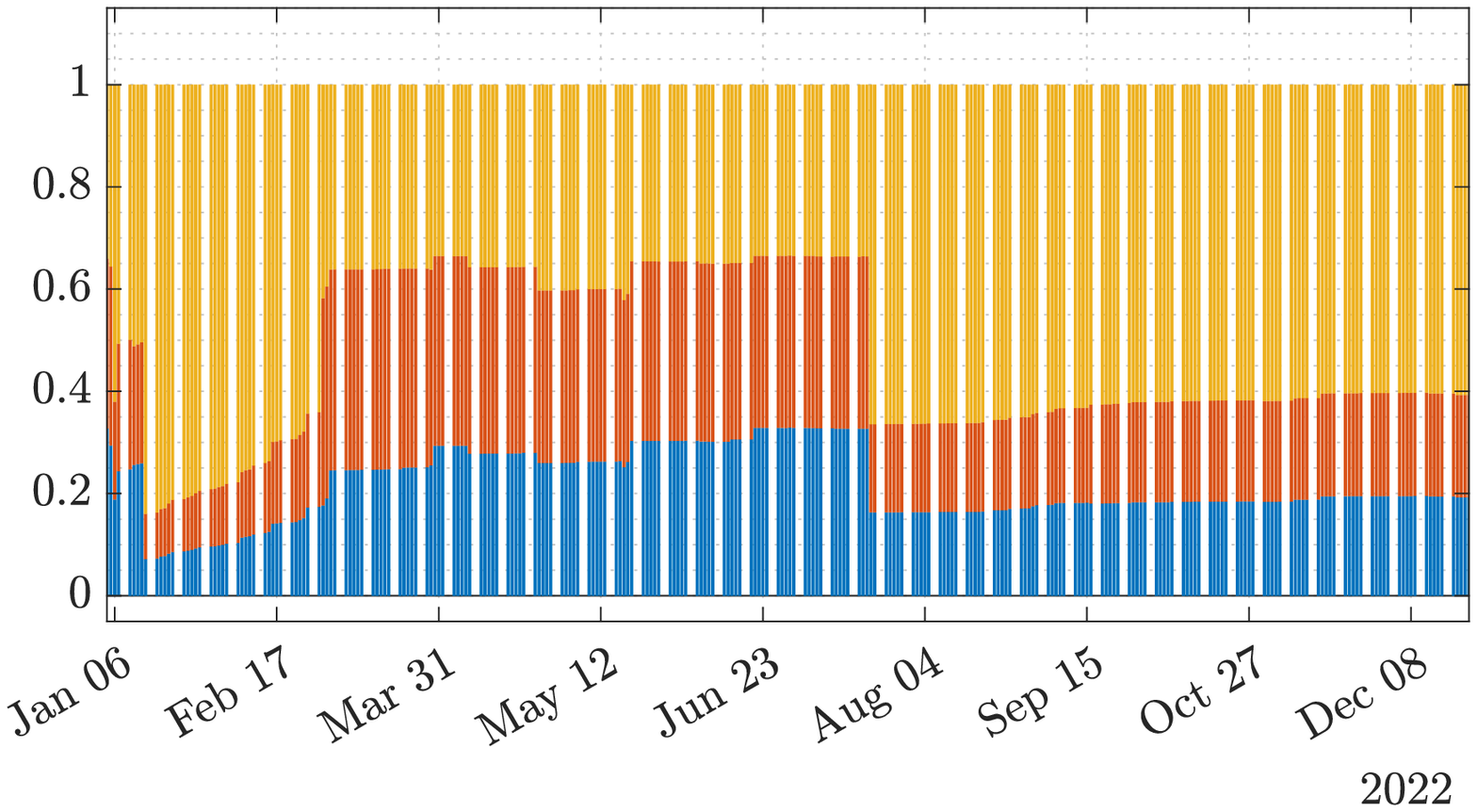} &
\includegraphics[trim= 0mm 0mm 0mm 0mm,clip, width= 5.0cm]{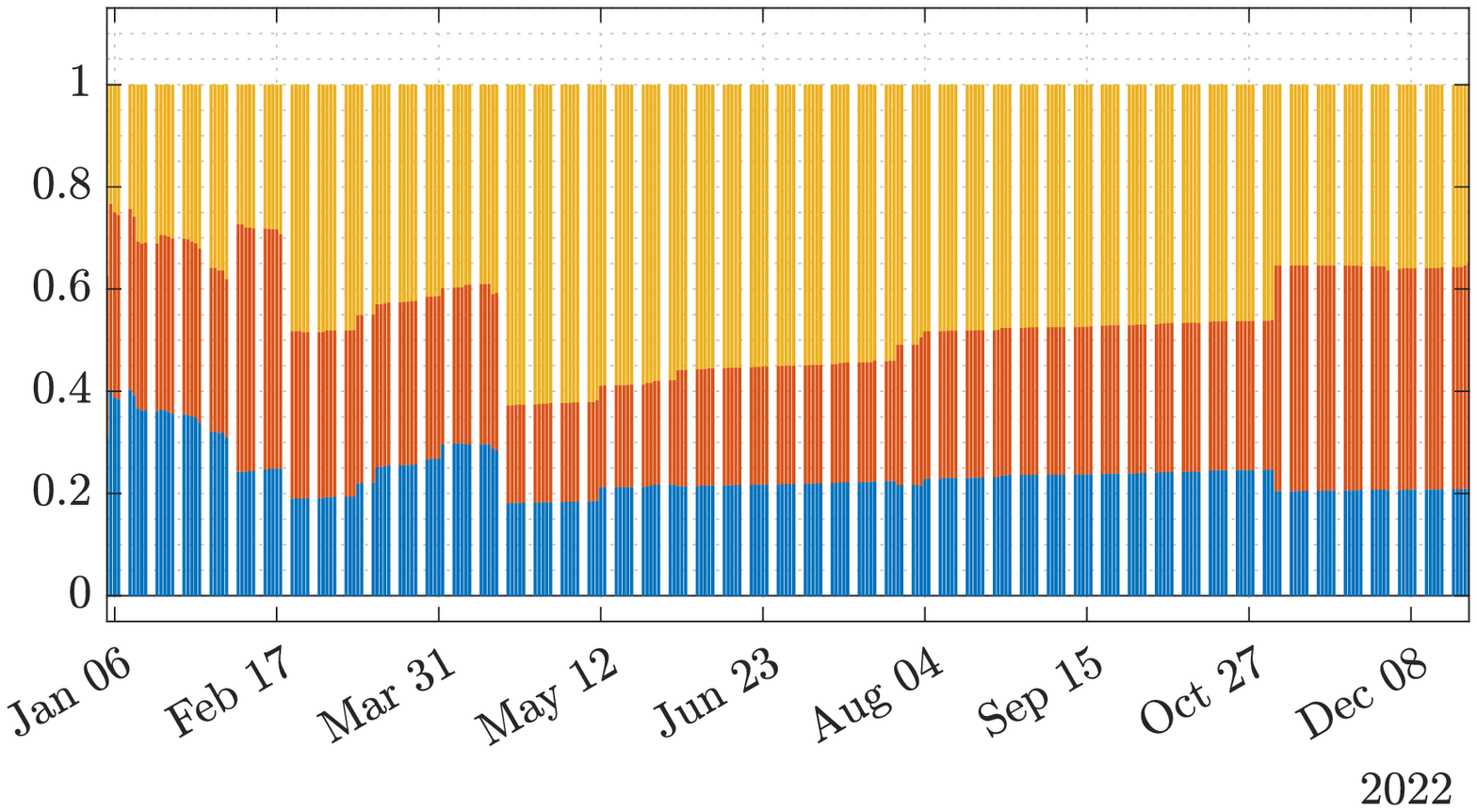} &
\includegraphics[trim= 0mm 0mm 0mm 0mm,clip, width= 5.0cm]{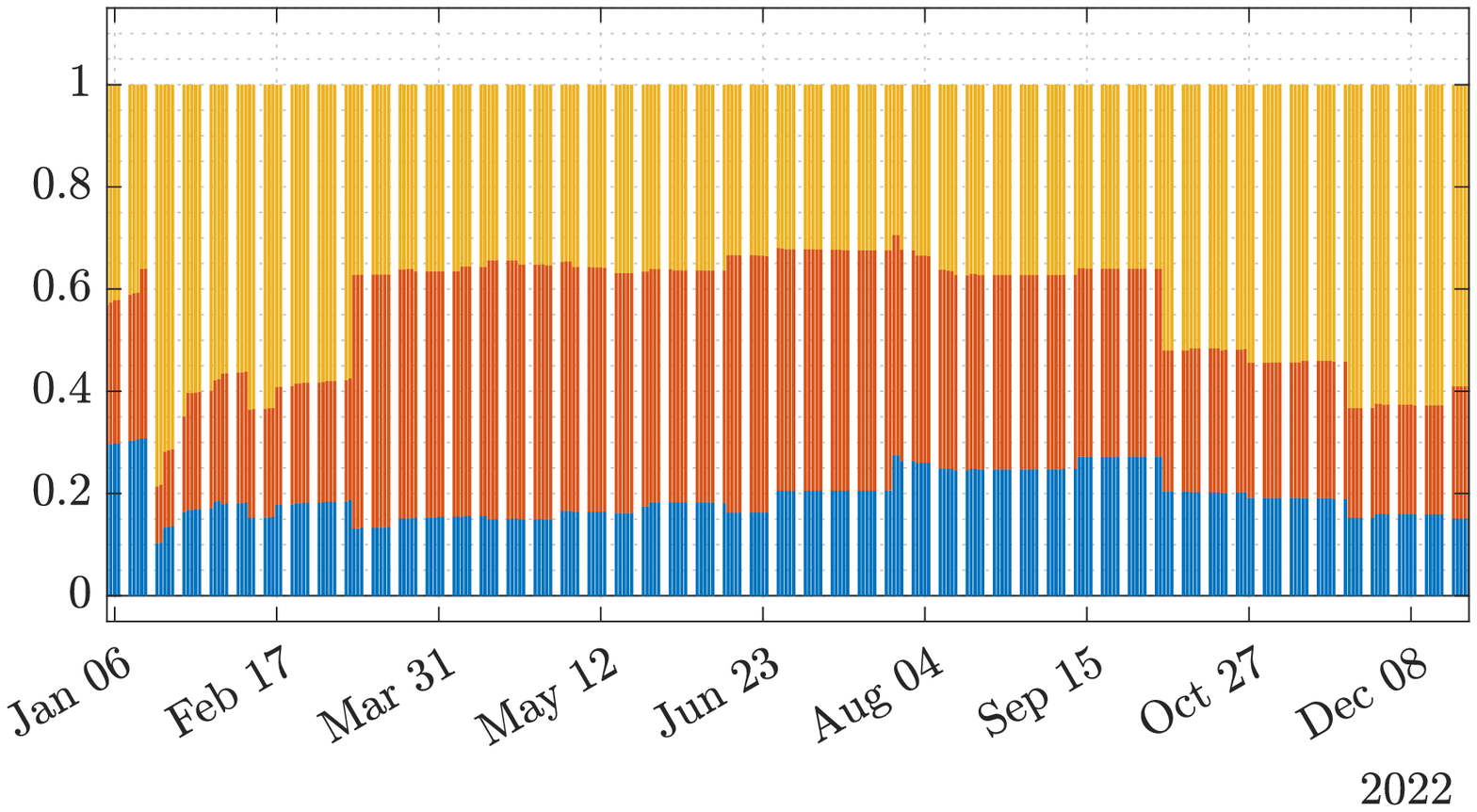} \\
\begin{rotate}{90} \hspace{30pt} {\scriptsize Coal} \end{rotate} \hspace*{-10pt} &
\includegraphics[trim= 0mm 0mm 0mm 0mm,clip, width= 5.0cm]{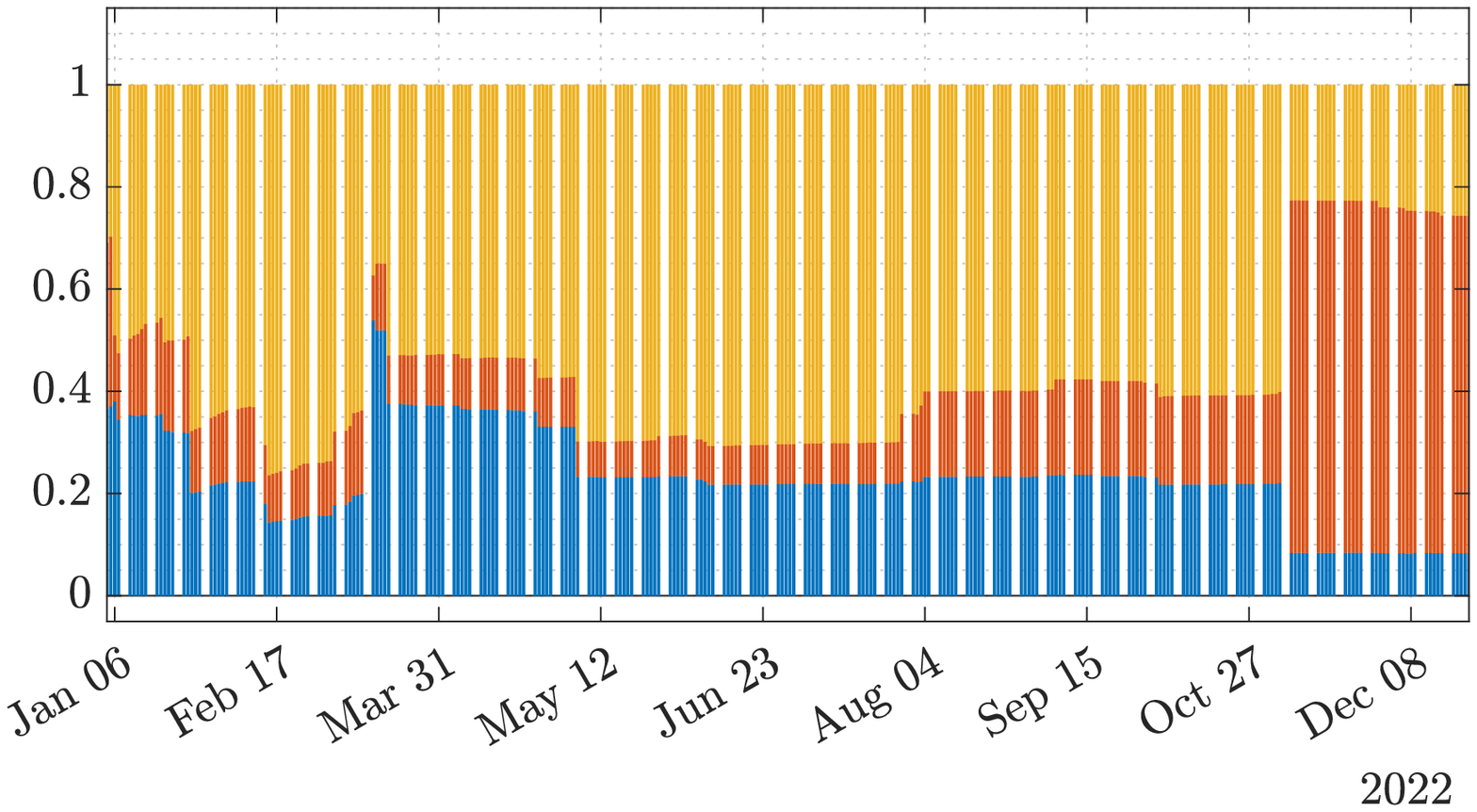} &
\includegraphics[trim= 0mm 0mm 0mm 0mm,clip, width= 5.0cm]{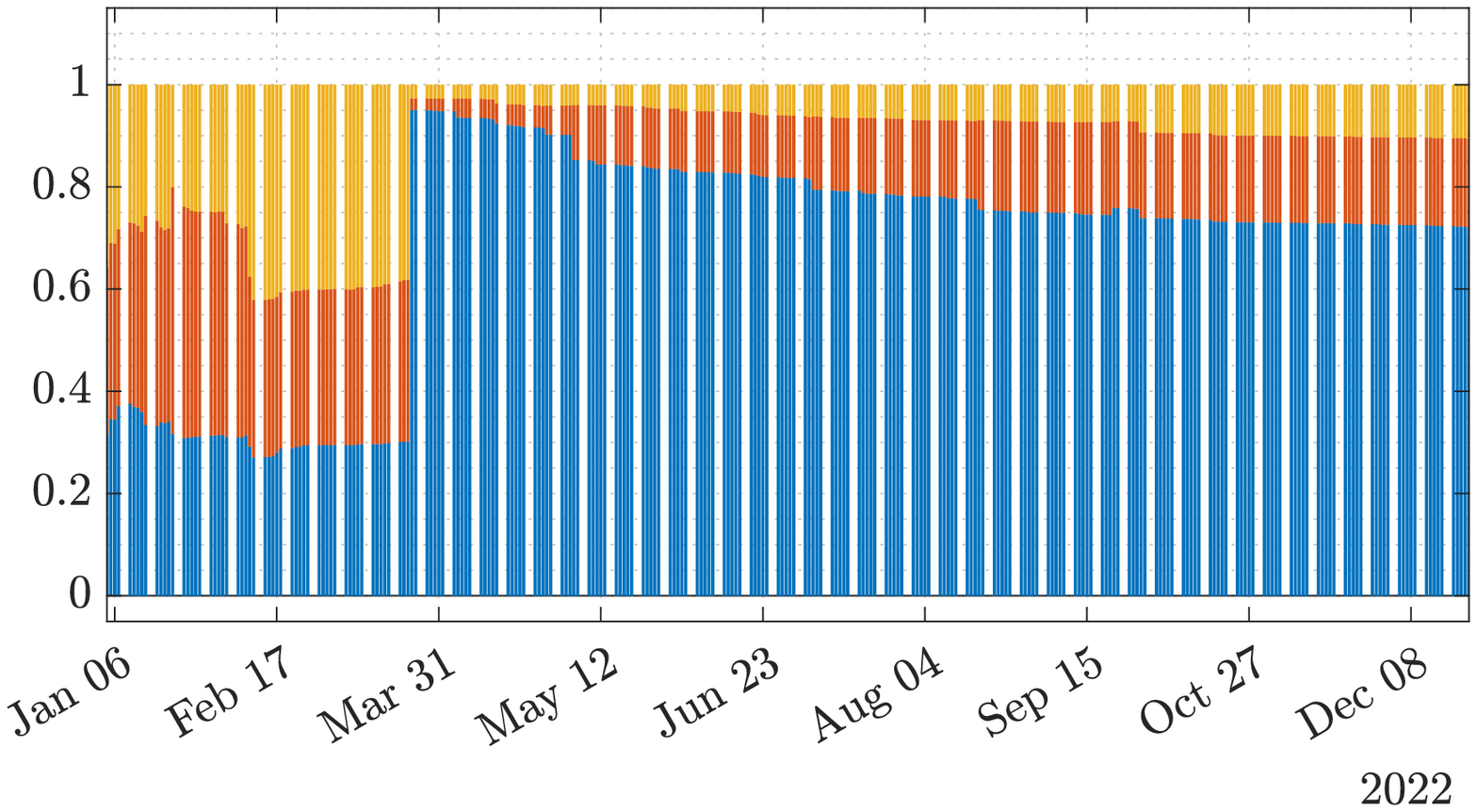} &
\includegraphics[trim= 0mm 0mm 0mm 0mm,clip, width= 5.0cm]{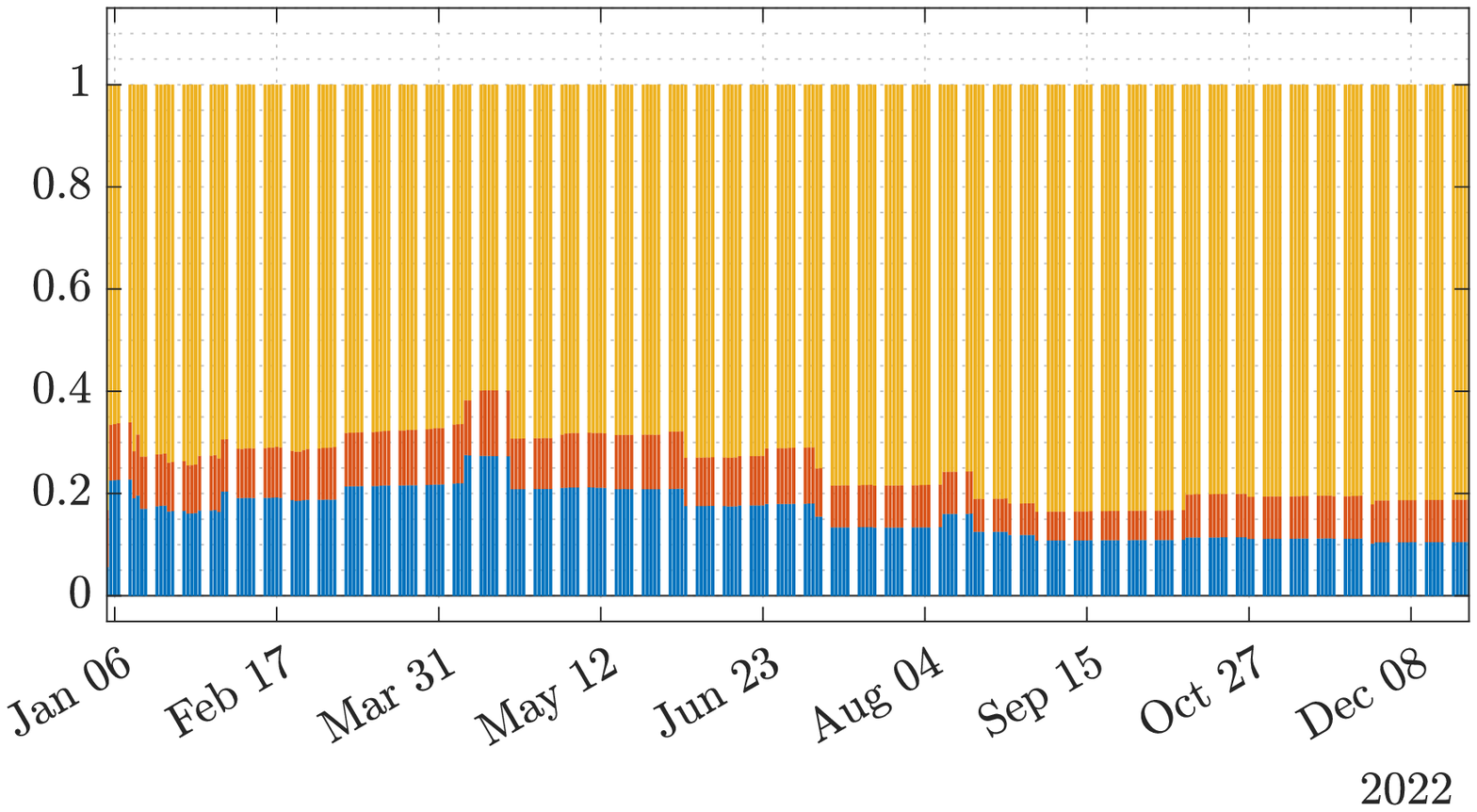} \\
\begin{rotate}{90} \hspace{30pt} {\scriptsize Gas} \end{rotate} \hspace*{-10pt} &
\includegraphics[trim= 0mm 0mm 0mm 0mm,clip, width= 5.0cm]
{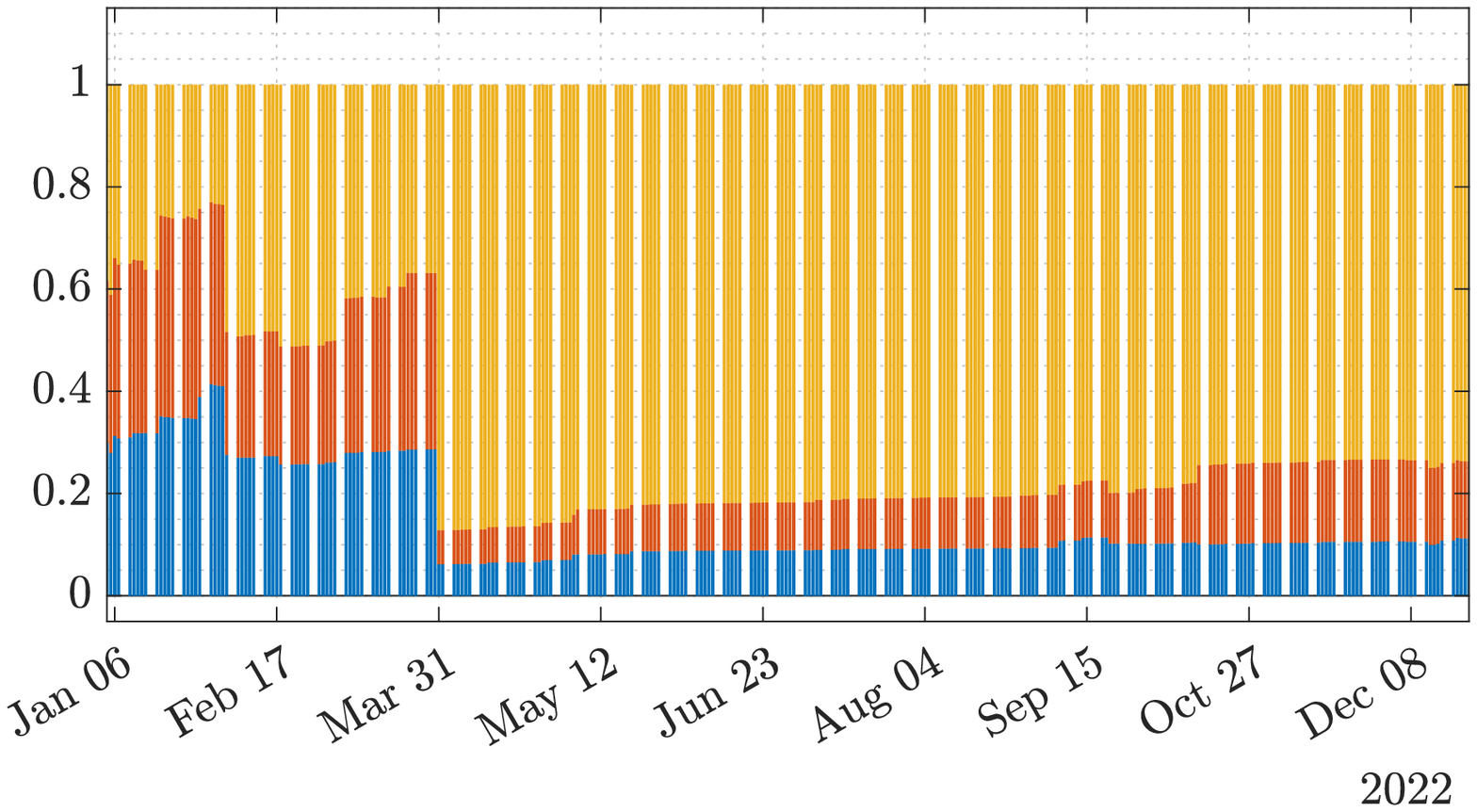} &
\includegraphics[trim= 0mm 0mm 0mm 0mm,clip, width= 5.0cm]{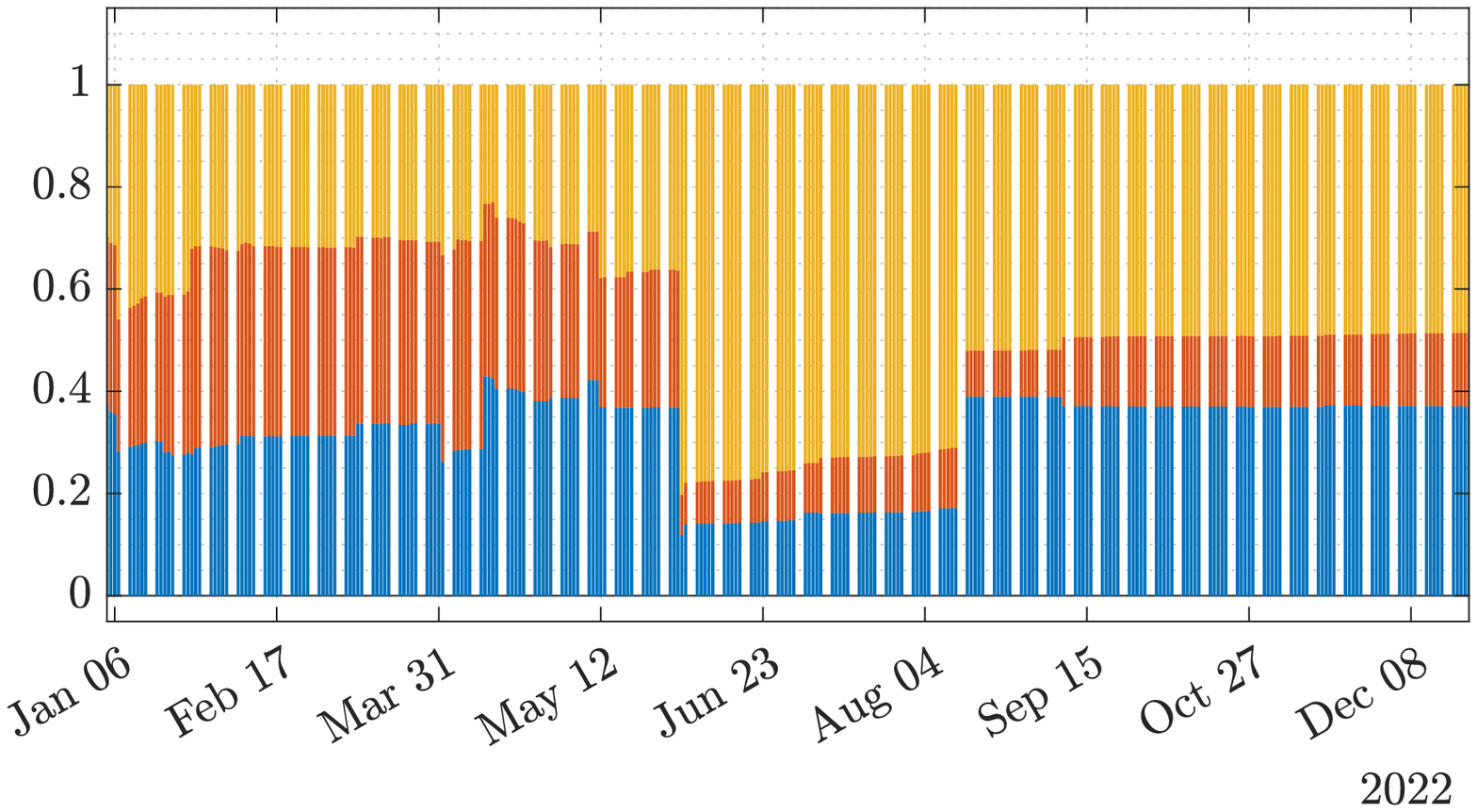} &
\includegraphics[trim= 0mm 0mm 0mm 0mm,clip, width= 5.0cm]{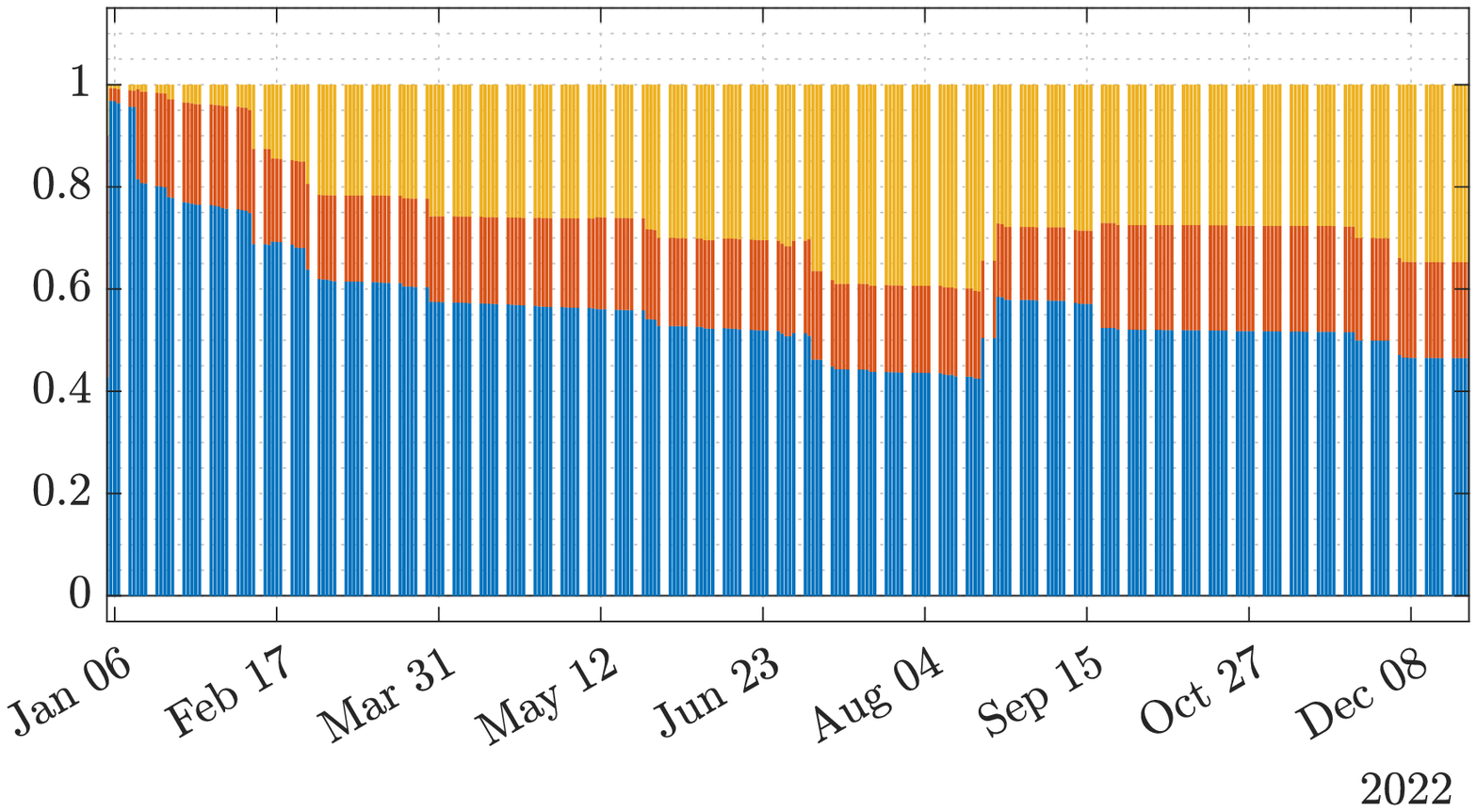}
\end{tabular}
\caption{Time-varying combination weights at horizon $h=1$ for Brent (first row), CO2 (second), Coal (third), and Gas (fourth) for the QVAR-SV (orange), QVAR-GARCH (yellow), and QVAR (blue) across different quantiles, $\tau = 0.1$ (left), $\tau = 0.5$ (centre), and $\tau = 0.9$ (right).}
\label{fig:Combin_roll_horiz1}
\end{figure}

\begin{figure}[t!h]
\centering
\hspace*{-2.8ex}
\begin{tabular}{c ccc}
 & {\small $\tau = 0.1$} & {\small $\tau = 0.5$} & {\small $\tau = 0.9$} \\
\begin{rotate}{90} \hspace{30pt} {\scriptsize Brent} \end{rotate} \hspace*{-10pt} &
\includegraphics[trim= 0mm 0mm 0mm 0mm,clip, width= 5.0cm]{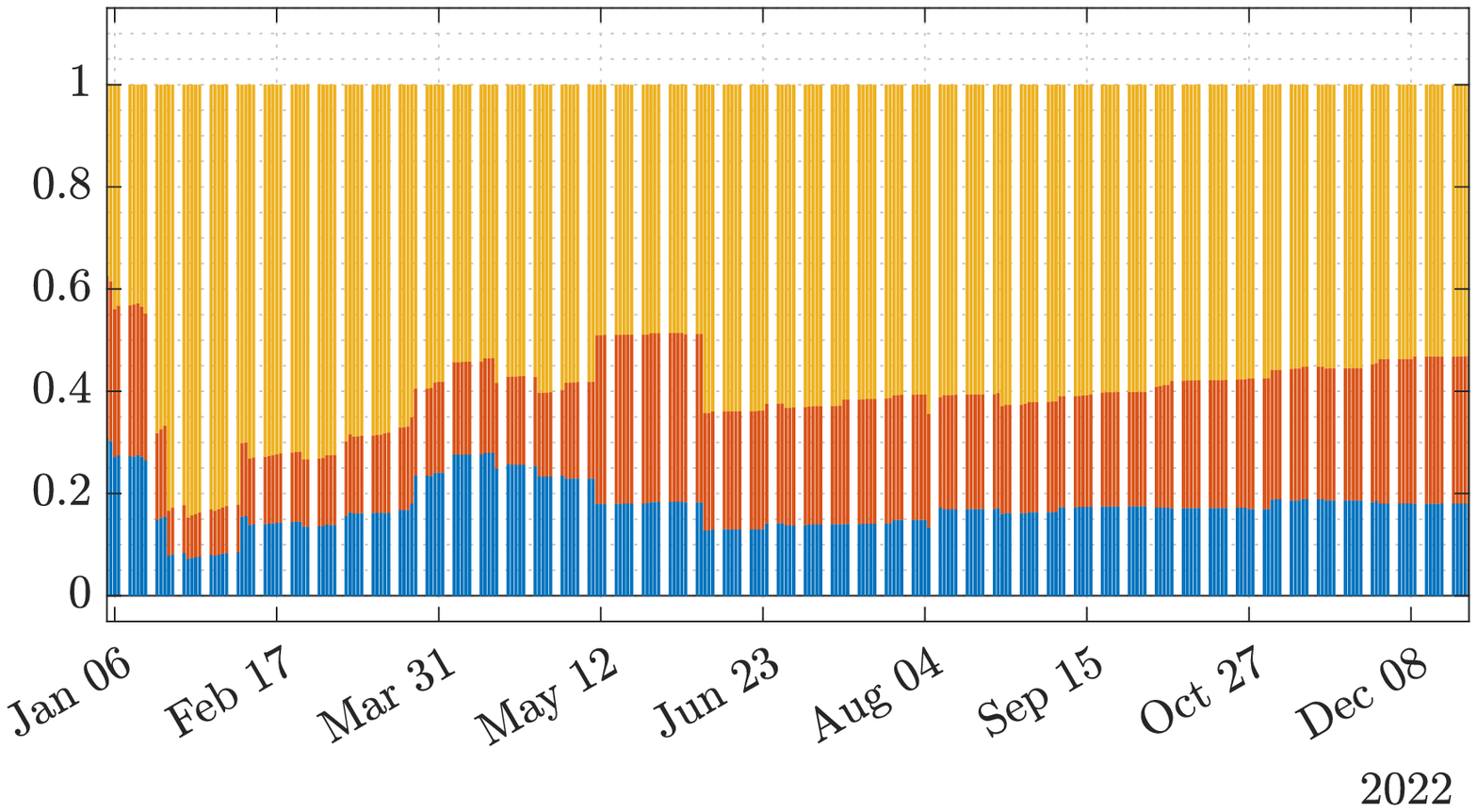} &
\includegraphics[trim= 0mm 0mm 0mm 0mm,clip, width= 5.0cm]{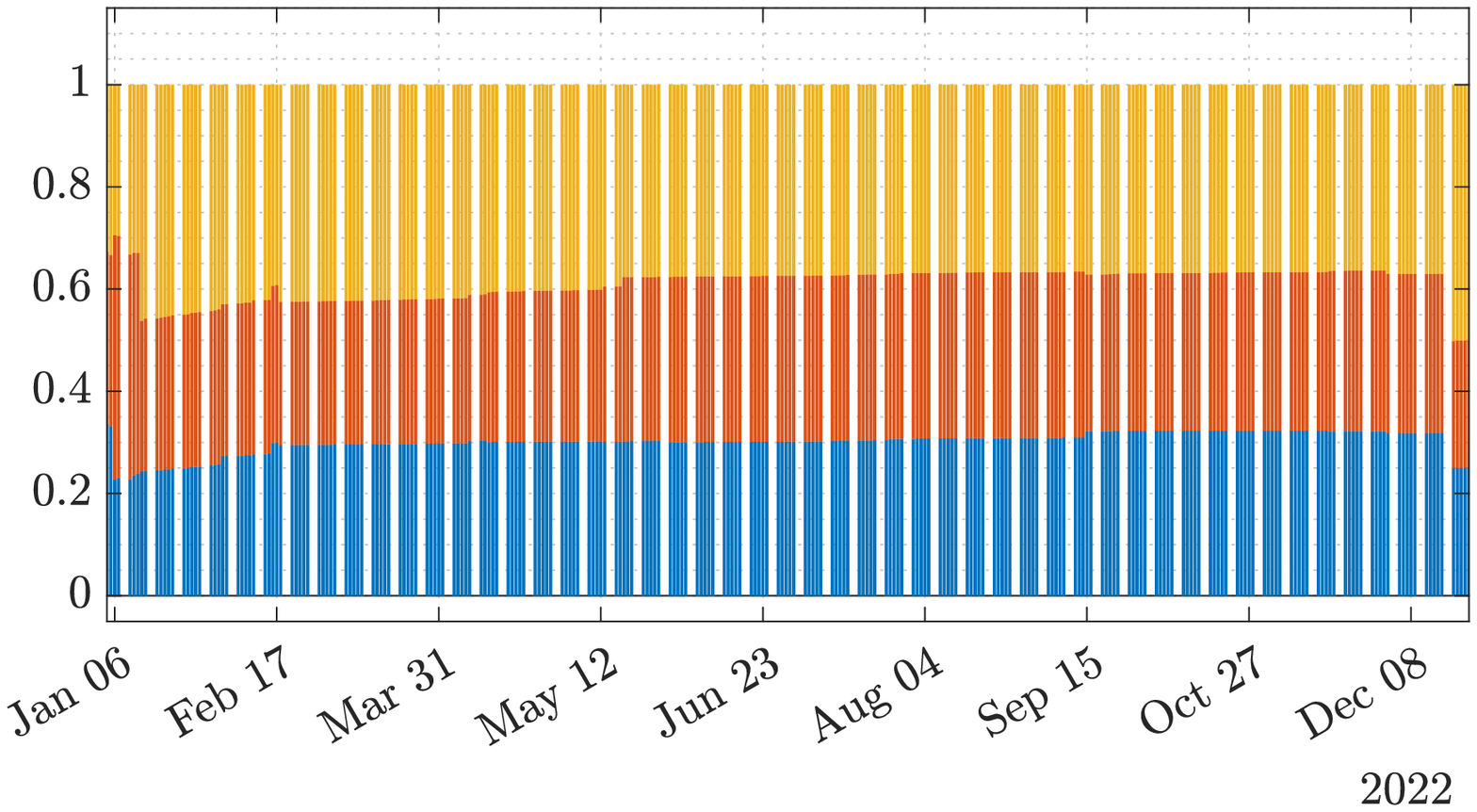} &
\includegraphics[trim= 0mm 0mm 0mm 0mm,clip, width= 5.0cm]{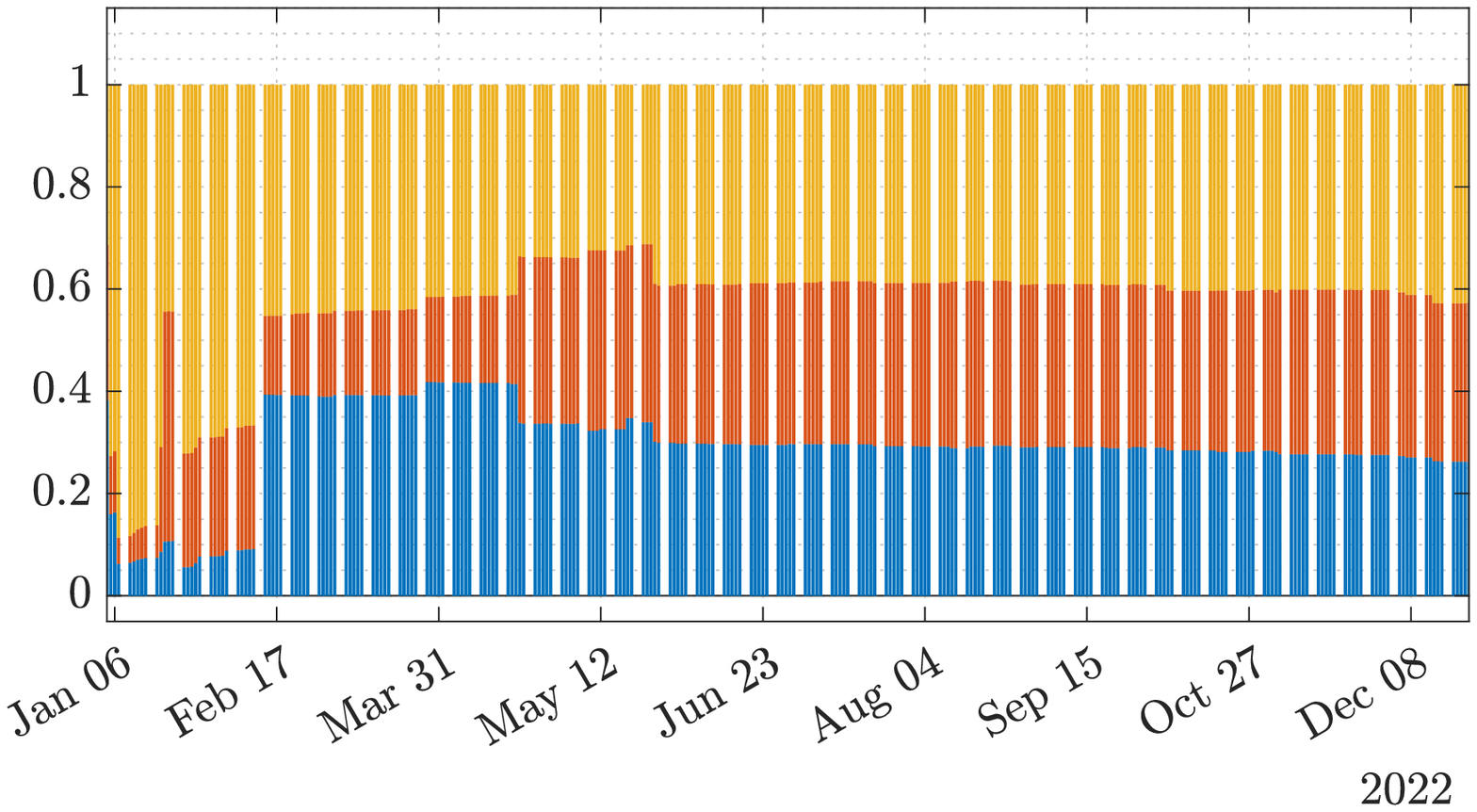} \\
\begin{rotate}{90} \hspace{30pt} {\scriptsize CO2} \end{rotate} \hspace*{-10pt} &
\includegraphics[trim= 0mm 0mm 0mm 0mm,clip, width= 5.0cm]{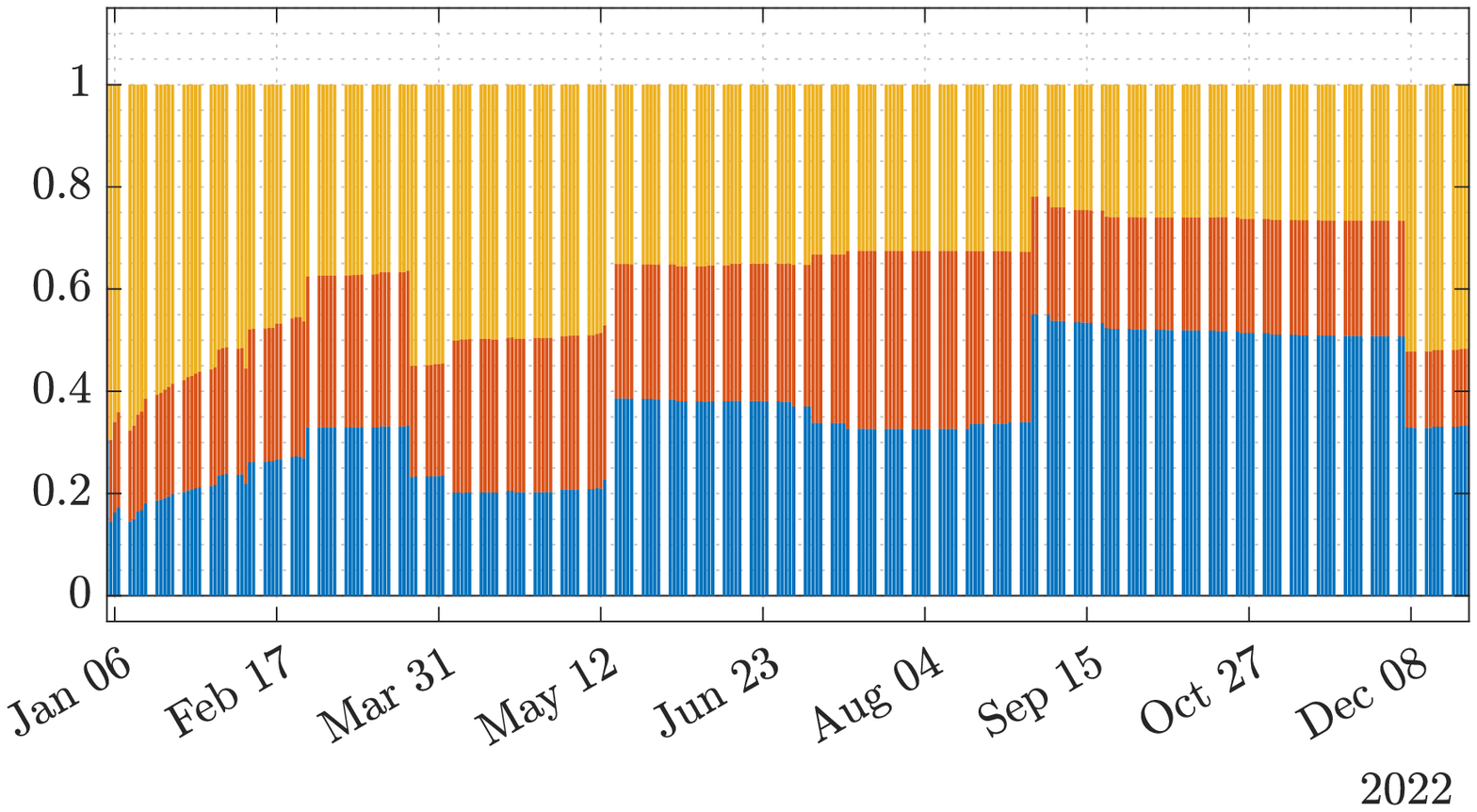} &
\includegraphics[trim= 0mm 0mm 0mm 0mm,clip, width= 5.0cm]{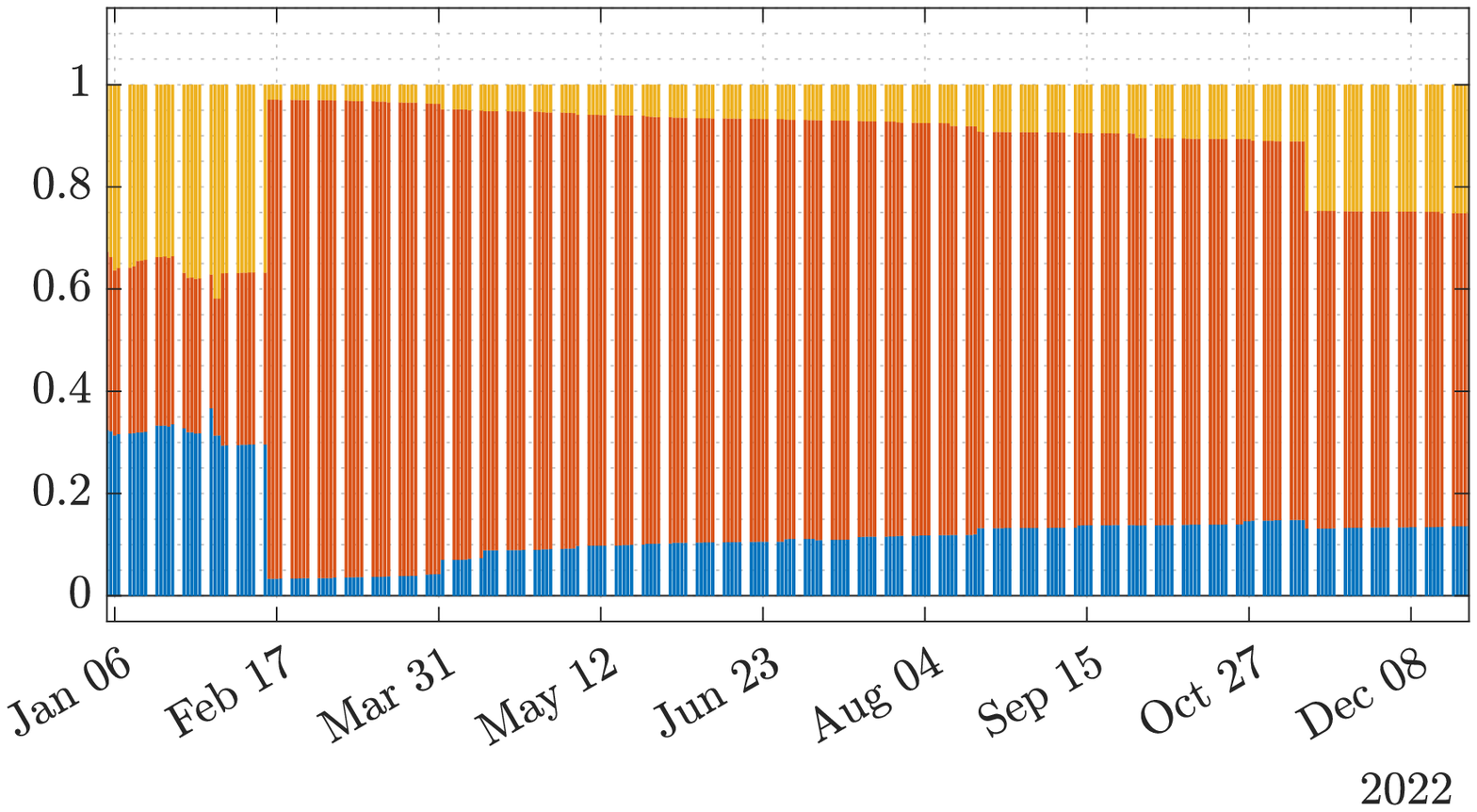} &
\includegraphics[trim= 0mm 0mm 0mm 0mm,clip, width= 5.0cm]{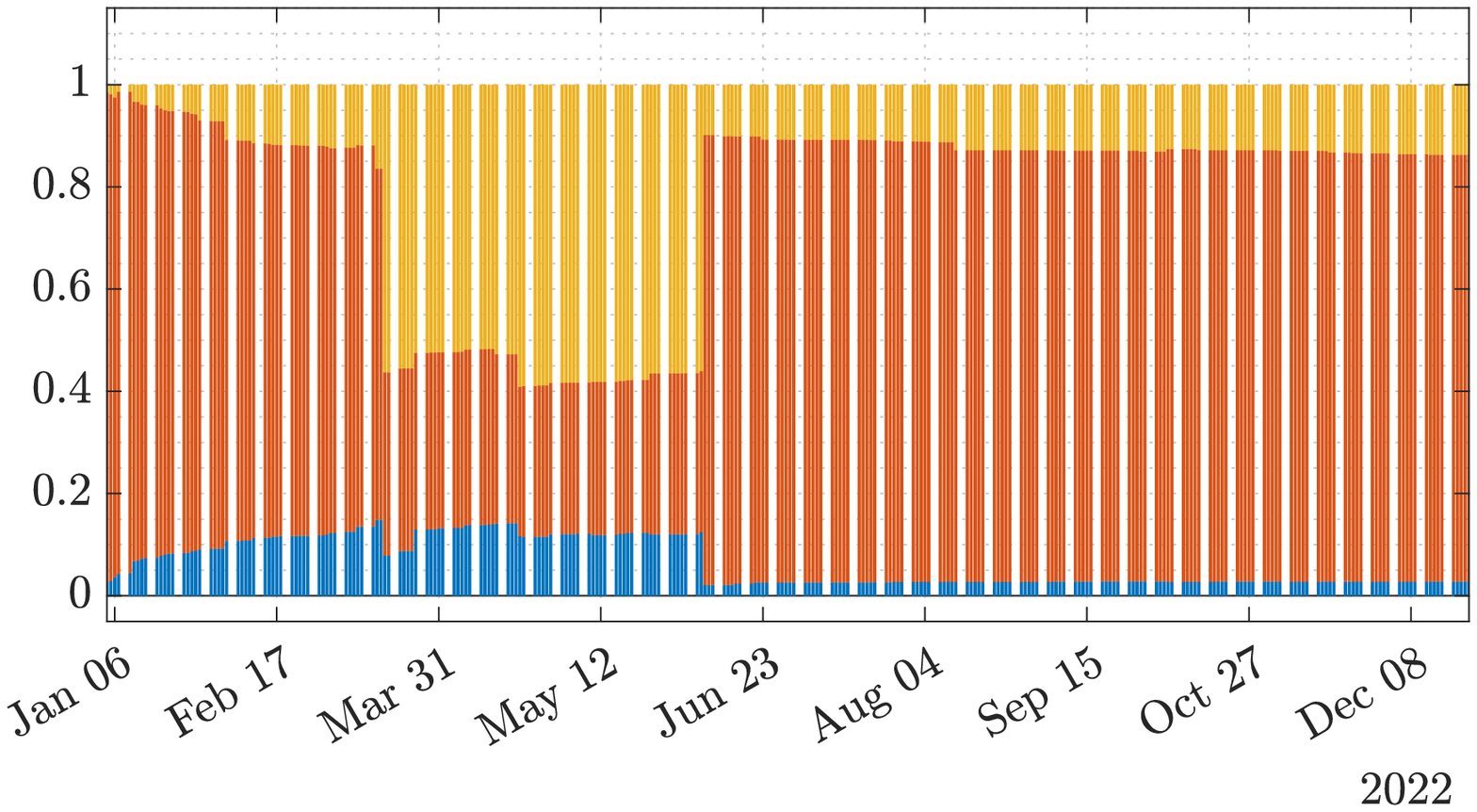} \\
\begin{rotate}{90} \hspace{30pt} {\scriptsize Coal} \end{rotate} \hspace*{-10pt} &
\includegraphics[trim= 0mm 0mm 0mm 0mm,clip, width= 5.0cm]{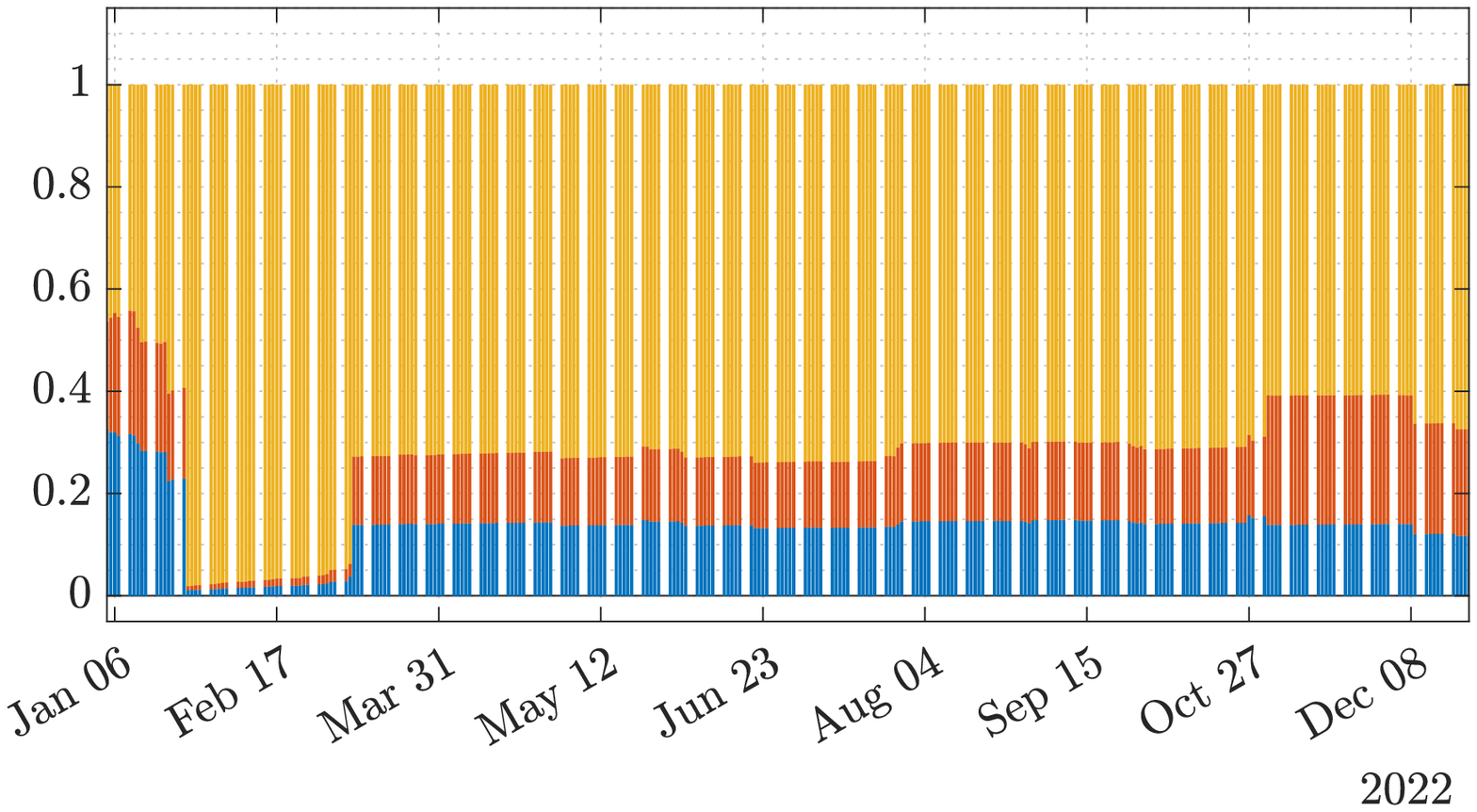} &
\includegraphics[trim= 0mm 0mm 0mm 0mm,clip, width= 5.0cm]{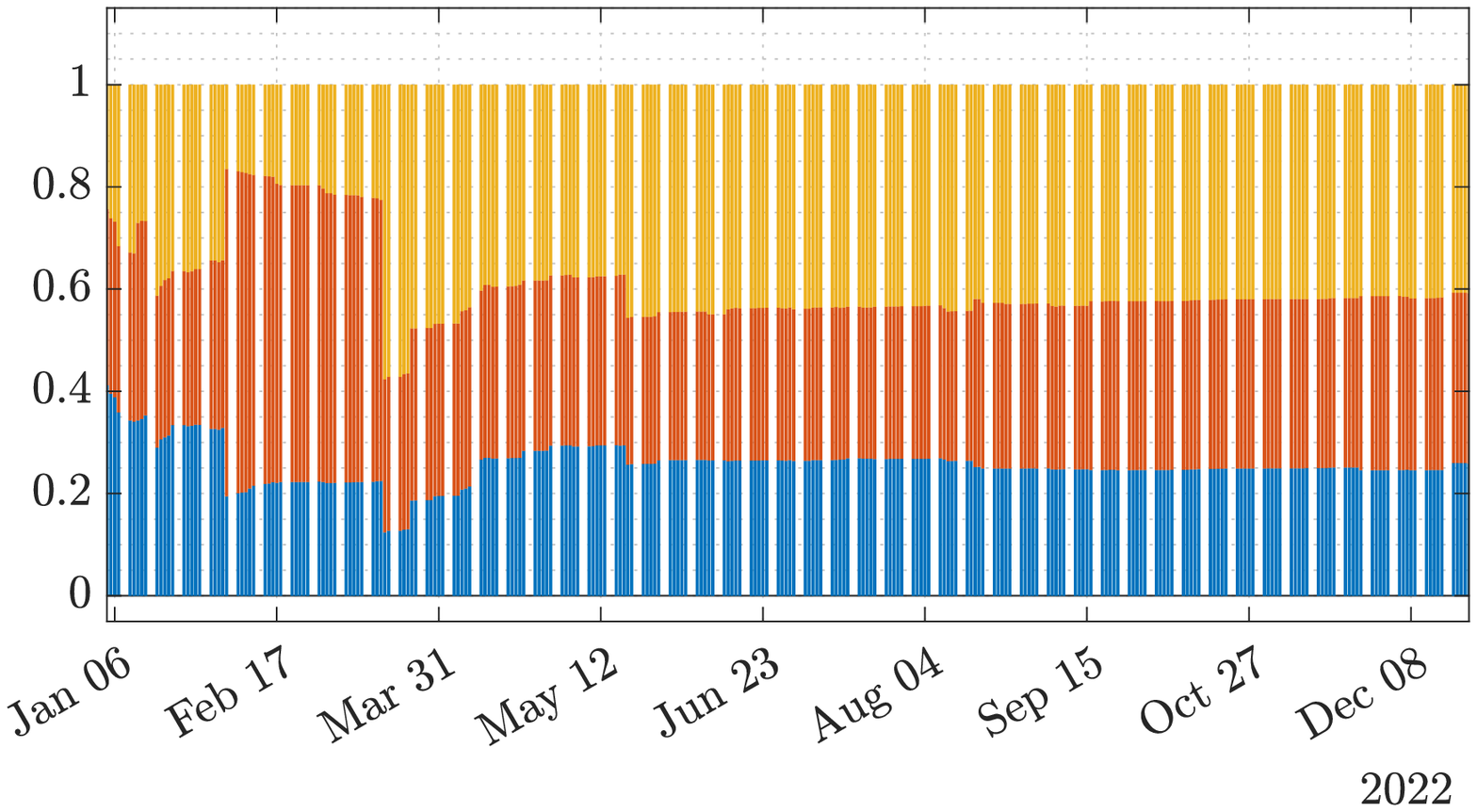} &
\includegraphics[trim= 0mm 0mm 0mm 0mm,clip, width= 5.0cm]{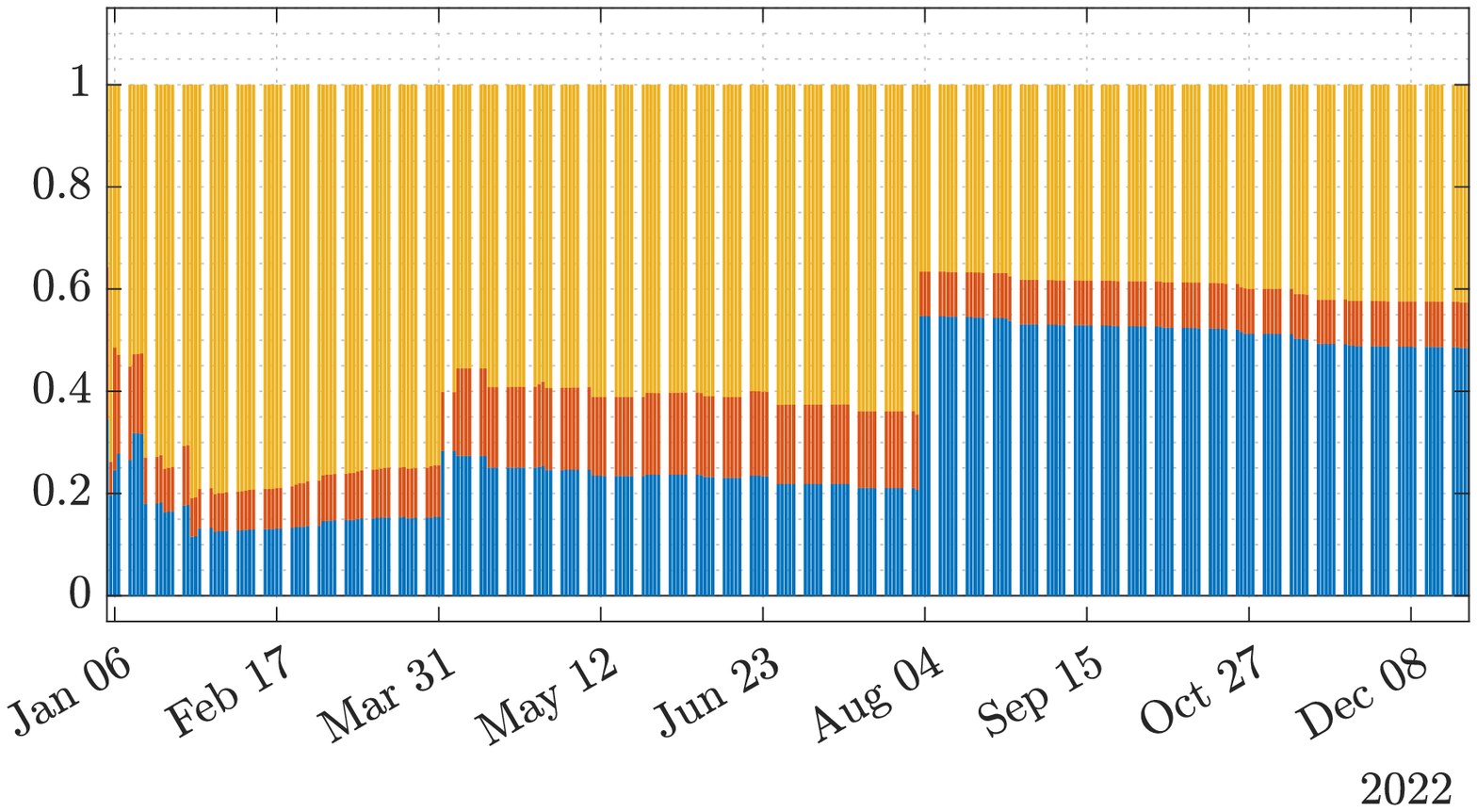} \\
\begin{rotate}{90} \hspace{30pt} {\scriptsize Gas} \end{rotate} \hspace*{-10pt} &
\includegraphics[trim= 0mm 0mm 0mm 0mm,clip, width= 5.0cm]
{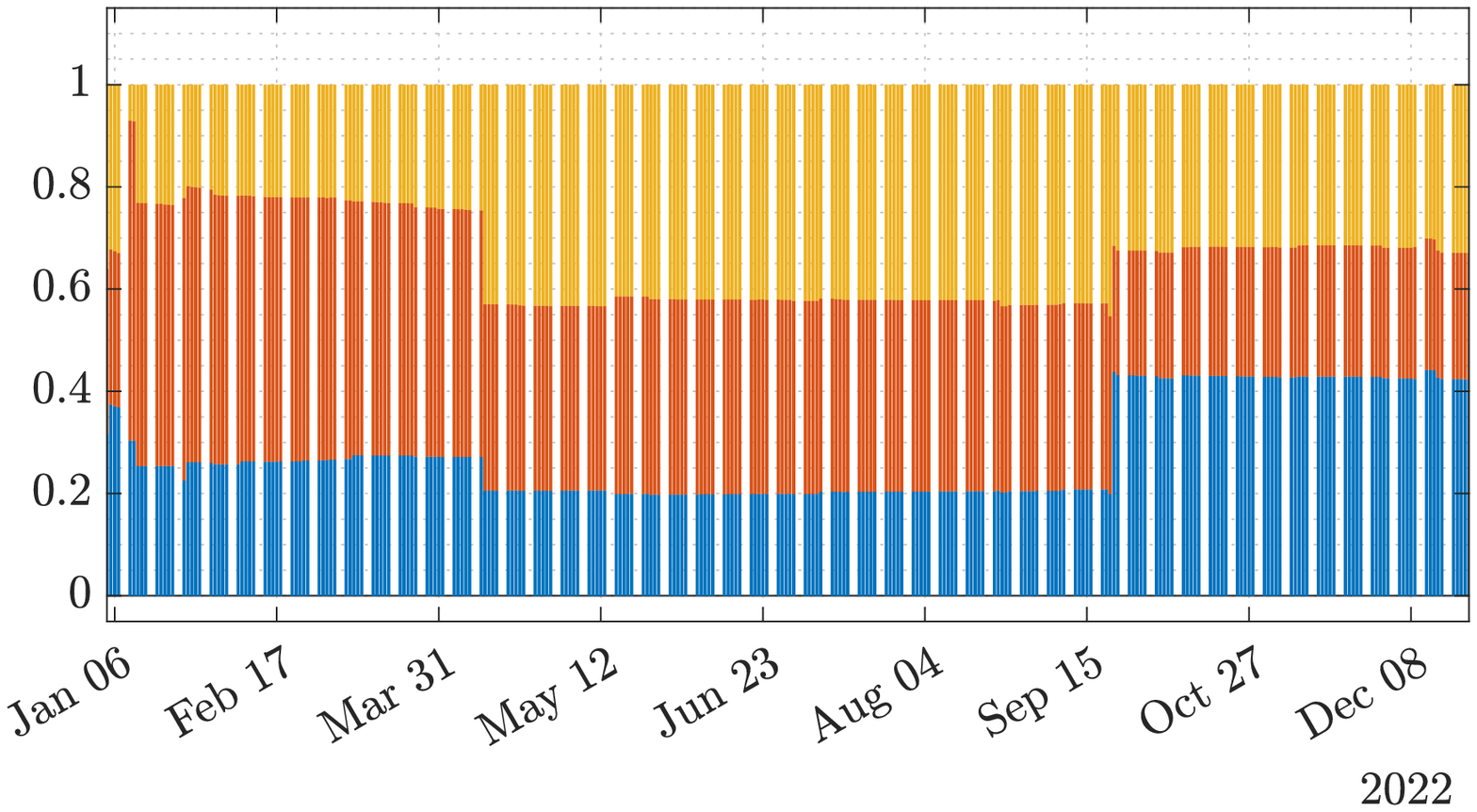} &
\includegraphics[trim= 0mm 0mm 0mm 0mm,clip, width= 5.0cm]{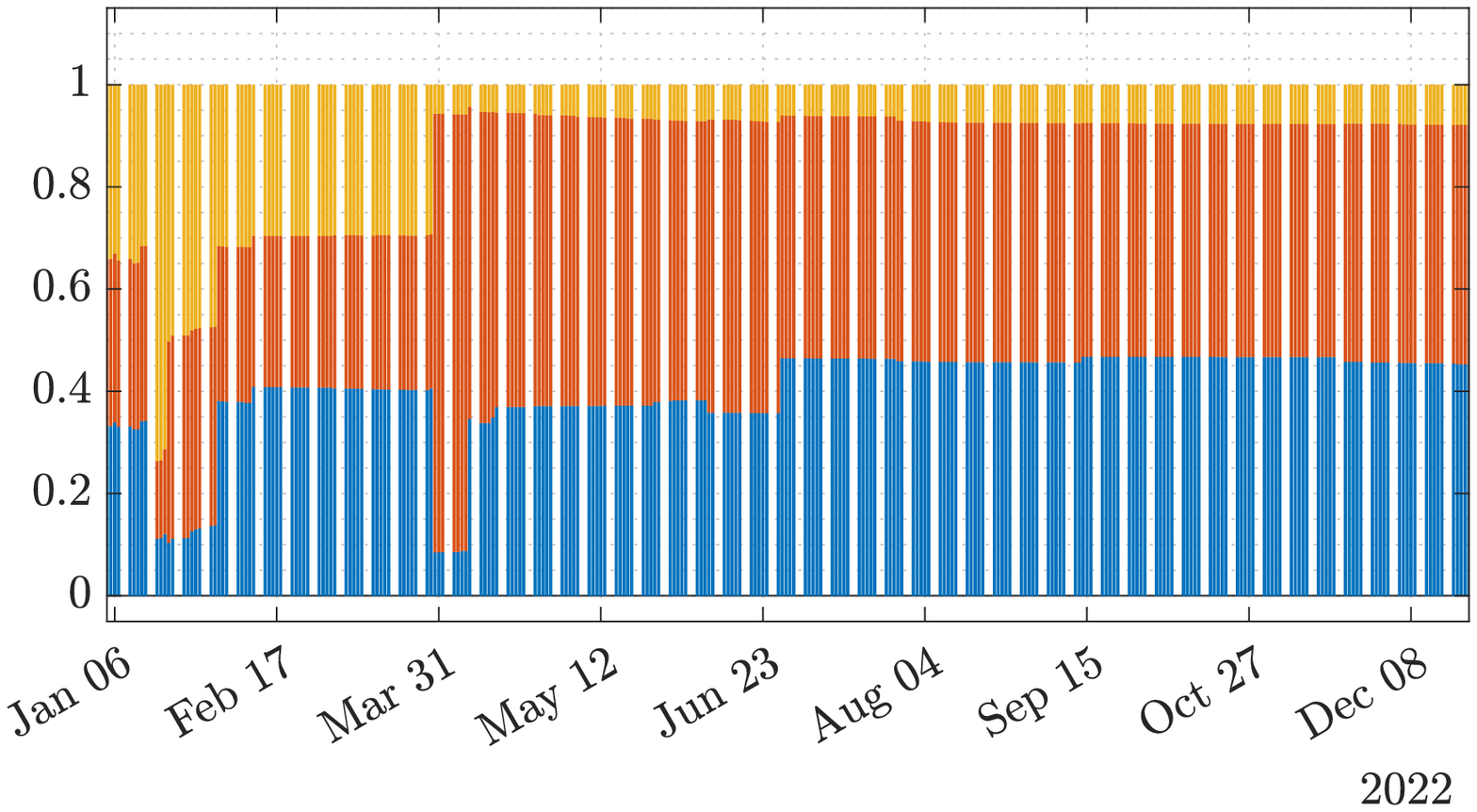} &
\includegraphics[trim= 0mm 0mm 0mm 0mm,clip, width= 5.0cm]{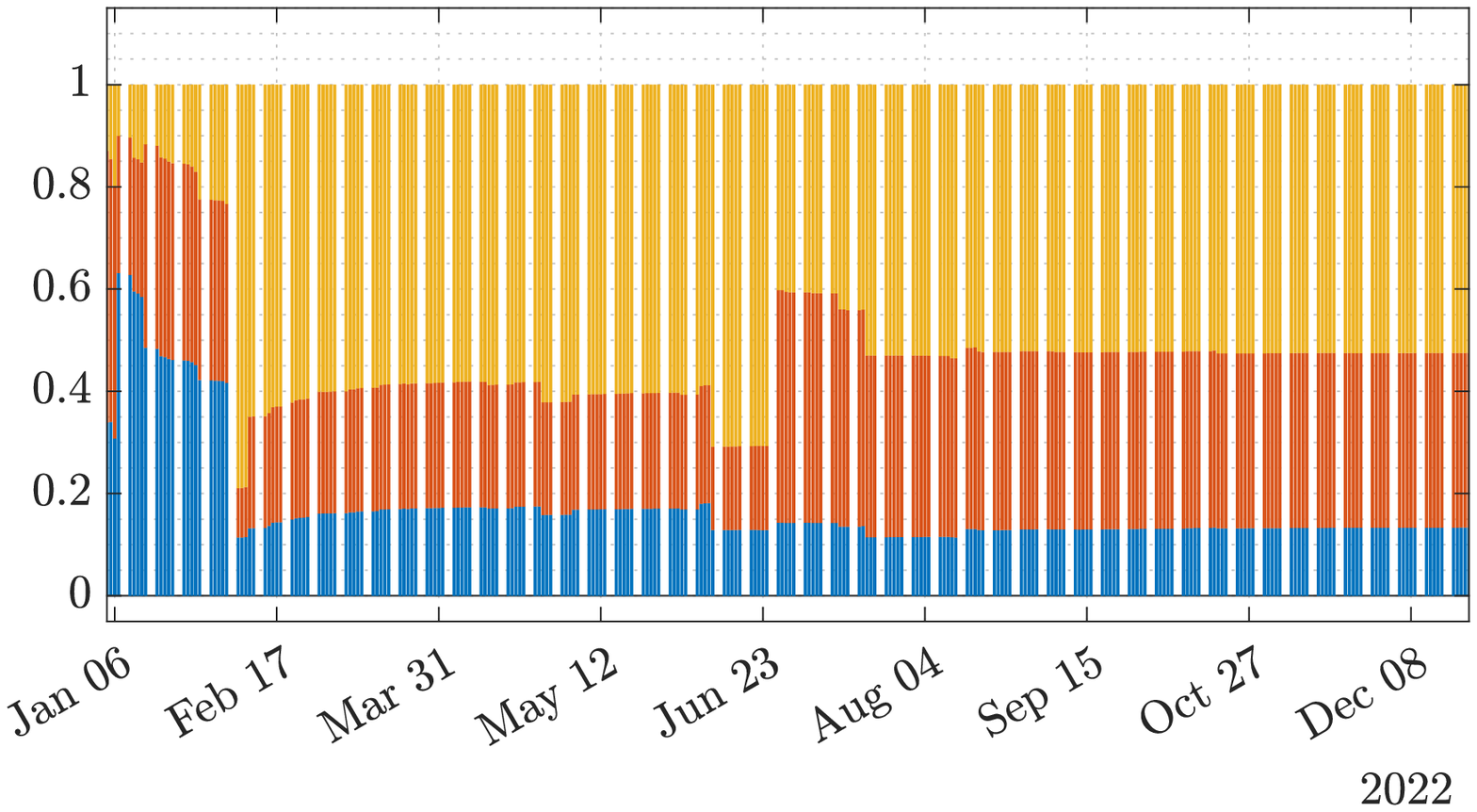}
\end{tabular}
\caption{Time-varying combination weights at horizon $h=5$ for Brent (first row), CO2 (second), Coal (third), and Gas (fourth) for the QVAR-SV (orange), QVAR-GARCH (yellow), and QVAR (blue) across different quantiles, $\tau = 0.1$ (left), $\tau = 0.5$ (centre), and $\tau = 0.9$ (right).}
\label{fig:Combin_roll_horiz5}
\end{figure}

Figure~\ref{fig:Combin_roll_horiz1} shows the time-varying weights at horizon $1$ associated with the three QVAR models with constant volatility, stochastic volatility, and GARCH effects for the four variables at the $10$th, $50$th, and $90$th percentiles.
The results show substantial variation in the model weights, both over time and cross-sectionally across the variables.
Second, we find evidence of different relative performances of the two time-varying volatility specifications over time for the left and right tails. Moreover, the constant volatility framework lags behind in most of the variables and quantiles investigated (as shown by the small blue bars), thus supporting the claim that paralleling conditional mean models, time-varying volatility improves the forecasting performance for conditional quantiles.
Finally, especially on the tails, the time-varying volatility models (especially the GARCH formulation) are found to greatly outperform the baseline QVAR, with cumulative weights between 70\% and 90\%. These patterns are less evident in the median quantile.

Moving to the time-varying weights at horizon $h=5$, in Fig.~\ref{fig:Combin_roll_horiz5} the importance of introducing time-varying models is more apparent, but there are different patterns with respect to horizon $h=1$. In particular, for both tails of the $CO_2$ we notice a strong preference for the QVAR-SV model. Similarly, both tails of Gas show that including time-varying volatility in the form of GARCH or stochastic volatility is preferable to a constant volatility model. In general, at a longer horizon, the time-varying volatility models outperform the constant volatility at any of the quantiles considered.

A possible explanation for the volatility of the combination weights is model misspecification; in particular, it might be that the unknown dynamical process for the scale might differ from the proposed SV and GARCH specifications. The model combination approach is naturally designed to address this issue, as demonstrated later in Table~\ref{tab:Result_rolling_horiz1}. However, we stress that for this dataset, the evidence supports the superiority of time-varying scale models compared to the constant baseline.

The Supplement reports the time-varying combination weights for all seventeen quantiles at both horizons. For each variable and quantile, we find results in line with the main takeaways presented in the paper. 
Overall, these findings support the claim that commonly used approaches modelling the conditional mean are likely to miss meaningful changes in the tail risks, thus supporting using a multivariate quantile regression framework.

{\tiny
\begin{ThreePartTable}
\begin{TableNotes}
\tiny{
\item[1] Please refer to Section \ref{sec:model} for the details on models. All forecasts are produced with a $h$-step-ahead rolling window process with a recursive approach.
\item[2] Real quantile scores are used for the QVAR model, and ratios for all the other models
\item[3] $^{\ast \ast \ast}$, $^{\ast \ast}$, $^{\ast}$ indicate that ratios are significantly different from 1 at $1\%$, $5\%$, $10\%$, according to the Diebold-Mariano test.
\item[4] Bold numbers indicate models belonging to the Superior Set of Models delivered by the \textsc{MCS} at confidence level $10\%$.
}
\end{TableNotes}
\captionsetup{width=0.95\textwidth}
\setlength{\tabcolsep}{3.2pt}
\begin{longtable}{l*{9}{c}}
\caption{\label{tab:Result_rolling_horiz1} Quantile Score for different variables (columns) and percentiles ($\tau = 0.1, 0.3, 0.5, 0.7$ and $0.9$) at horizon $1$.}\\
\hline  %\\[-0.4cm]
  Variable      & Brent & $CO_2$ & Coal & Gas & & Brent & $CO_2$ & Coal & Gas \\[0.01mm]
\hline  %\\[-0.4cm]
  & \multicolumn{4}{c}{\textbf{Quantile: $\tau = 0.10$}} & & \multicolumn{4}{c}{\textbf{Quantile: $\tau = 0.30$}}\\
QVAR       & 1.148 & 1.009 & 0.879 & 0.977 & & 0.612 & 0.544 & 0.436 & 0.559 \\
QVAR-SV    & \bf0.838\threestar & \bf0.928\threestar & 1.103 & 0.941\threestar & & \bf0.964\threestar & \bf0.973\threestar & 1.094 & 0.981\threestar \\
QVAR-GARCH & 0.907\threestar & \bf0.913\threestar & \bf0.769\threestar & \bf0.862\threestar & & \bf0.966\threestar & \bf0.945\threestar & \bf0.868\threestar & \bf0.936\threestar \\
QVAR Combination (AVG) & 0.894\threestar & 0.932\threestar & 0.928\threestar & 0.881\threestar & & 0.972\threestar & \bf0.957\threestar & 0.957\threestar & 0.954\threestar \\
QVAR Combination (T-V) & 0.893\threestar & 0.933\threestar & 0.885\threestar & \bf0.879\threestar & & 0.974\threestar & \bf0.957\threestar & 0.957\threestar & 0.957\threestar \\
  & \multicolumn{4}{c}{\textbf{Quantile: $\tau = 0.50$}} & & \multicolumn{4}{c}{\textbf{Quantile: $\tau = 0.70$}}\\
QVAR       & \bf0.453 & \bf0.381 & \bf0.259 & \bf0.400 & & 0.600 & 0.564 & 0.437 & \bf0.560 \\
QVAR-SV    & \bf0.997 & \bf0.999 & \bf1.003 & \bf0.999 & & \bf0.957\threestar & \bf0.977\threestar & 1.085 & \bf0.988\threestar \\
QVAR-GARCH & \bf0.994 & \bf1.002 & \bf1.020 & \bf0.992 & & 1.013 & \bf0.958\twostar & \bf0.888\threestar & \bf1.003 \\
QVAR Combination (AVG) & \bf0.995\onestar & \bf0.997 & \bf0.999 & \bf0.989\onestar & & 0.992 & 0.987\threestar & 0.914\threestar & \bf0.985\threestar \\
QVAR Combination (T-V) & \bf0.995\onestar & \bf0.996 & \bf0.996 & \bf0.993 & & 0.994 & 0.984\threestar & 0.920\threestar & \bf0.987\twostar \\
  & \multicolumn{4}{c}{\textbf{Quantile: $\tau = 0.90$}} & \\
QVAR       & 1.155 & 1.069 & 0.882 & 1.022 \\
QVAR-SV    & \bf0.824\threestar & 0.922\threestar & 1.091 & 0.937\threestar \\
QVAR-GARCH & 0.912\threestar & \bf0.841\threestar & \bf0.739\threestar & \bf0.872\threestar \\
QVAR Combination (AVG) & 0.864\threestar & 0.896\threestar & 0.798\threestar & 0.948\threestar \\
QVAR Combination (T-V) & 0.867\threestar & 0.899\threestar & 0.801\threestar & 0.940\threestar \\
\hline 
\insertTableNotes  % tell LaTeX where to insert the contents of "TableNotes"
\end{longtable}
\end{ThreePartTable}
}

{\tiny
\begin{ThreePartTable}
\begin{TableNotes}
\tiny{
\item[] Please see the notes to Table~\ref{tab:Result_rolling_horiz1}.
%\item[1] Please refer to Section \ref{sec:model} for the details on models. All forecasts are produced with a $h$-step-ahead rolling window process with a recursive approach.
%\item[2] $^{\ast \ast \ast}$, $^{\ast \ast}$, $^{\ast}$ indicate that ratios are significantly different from 1 at $1\%$, $5\%$, $10\%$, according to the Diebold-Mariano test.
%\item[3] Bold numbers indicate models belonging to the Superior Set of Models delivered by the \textsc{MCS} at confidence level $10\%$.
}
\end{TableNotes}
\captionsetup{width=0.95\textwidth}
\setlength{\tabcolsep}{3.2pt}
\begin{longtable}{l*{9}{c}}
\caption{\label{tab:Result_rolling_horiz5} Quantile Score for different variables (columns) and percentiles ($\tau = 0.1, 0.3, 0.5, 0.7$ and $0.9$) at horizon $5$.}\\
\hline  %\\[-0.4cm]
  Variable      & Brent & $CO_2$ & Coal & Gas & & Brent & $CO_2$ & Coal & Gas \\[0.01mm]
\hline  %\\[-0.4cm]
  & \multicolumn{4}{c}{\textbf{Quantile: $\tau = 0.10$}} & & \multicolumn{4}{c}{\textbf{Quantile: $\tau = 0.30$}}\\
QVAR       & 1.264 & 0.930 & 1.024 & 1.127 & & 0.627 & 0.499 & 0.460 & \bf0.578 \\
QVAR-SV    & 0.831\threestar & \bf0.908\threestar & 1.067 & 0.937\threestar & & \bf0.964\threestar & \bf0.968\threestar & 1.078 & \bf0.986\threestar \\
QVAR-GARCH & \bf0.769\threestar & \bf0.926\threestar & \bf0.615\threestar & \bf0.775\threestar & & \bf0.958\threestar & 1.022 & \bf0.919\threestar & \bf1.005 \\
QVAR Combination (AVG) & 0.820\threestar & 0.953\threestar & 0.734\threestar & 0.897\threestar & & 0.992\threestar & 1.003 & 0.971\twostar & \bf1.000 \\
QVAR Combination (T-V) & 0.816\threestar & 0.937\threestar & 0.728\threestar & 0.890\threestar & & 0.986\threestar & 0.999 & \bf0.957\threestar & \bf0.997 \\
  & \multicolumn{4}{c}{\textbf{Quantile: $\tau = 0.50$}} & & \multicolumn{4}{c}{\textbf{Quantile: $\tau = 0.70$}}\\
QVAR       & \bf0.453 & \bf0.370 & \bf0.252 & \bf0.388 & & \bf0.614 & 0.526 & \bf0.457 & \bf0.589 \\
QVAR-SV    & \bf0.998 & \bf1.001 & \bf1.002 & \bf1.003 & & \bf0.958\threestar & \bf0.959\threestar & \bf1.064 & \bf0.988\threestar \\
QVAR-GARCH & \bf1.003 & \bf0.999 & \bf1.020 & 1.028 & & \bf1.008 & 1.032 & \bf1.063 & \bf1.108 \\
QVAR Combination (AVG) & \bf1.000 & \bf1.000 & \bf0.998 & \bf1.002 & & \bf0.990\onestar & 0.996\onestar & \bf1.029 & \bf1.029 \\
QVAR Combination (T-V) & \bf1.000 & \bf1.000 & \bf0.998 & \bf1.004 & & \bf0.985\twostar & 1.000 & \bf1.023 & \bf1.038 \\
  & \multicolumn{4}{c}{\textbf{Quantile: $\tau = 0.90$}} & \\
QVAR       & 1.226 & 0.960 & 0.965 & 1.114 \\
QVAR-SV    & 0.830\threestar & \bf0.898\threestar & 1.059 & 0.957\threestar \\
QVAR-GARCH & \bf0.775\threestar & \bf0.876\threestar & \bf0.599\threestar & \bf0.815\threestar \\
QVAR Combination (AVG) & 0.854\threestar & \bf0.897\threestar & 0.811\threestar & 0.884\threestar \\
QVAR Combination (T-V) & 0.851\threestar & 0.901\threestar & 0.806\threestar & 0.882\threestar \\
\hline 
\insertTableNotes  % tell LaTeX where to insert the contents of "TableNotes"
\end{longtable}
\end{ThreePartTable}
}

% In the out-of-sample forecasting exercise, we compare the QVAR model with $1$ lag and different time-varying volatility for various quantile levels $\tau \in \{ 0.1,0.15,\ldots,0.9 \}$ by using a rolling window approach.
Besides, we have also considered the forecast combination with either time-varying weights or time-averaged weights (see Section 3).
Tables~\ref{tab:Result_rolling_horiz1} and \ref{tab:Result_rolling_horiz5} show the results of the forecasting exercise across variables (in column) for five quantiles, $\tau \in \{ 0.1, 0.3, 0.5, 0.7, 0.9 \}$, using a rolling window process with a recursive approach for $1$- and $5$-steps ahead, respectively.
The tables also report the outcome of the Diebold-Mariano (DM) test against the benchmark (QVAR with constant volatility) and the model confidence set (MCS) at $10\%$ compared to all the models. Both tables provide the quantile scores for the QVAR model, while for the other models we report the ratio against the benchmark model, where entries lower than $1$ means that the model considered outperforms the QVAR model.

In line with the previous plots, Tab.~\ref{tab:Result_rolling_horiz1} provides evidence supporting the heterogeneity of the models' forecasting performance across the variables and the quantile investigated.
At the $10$th percentile, the QVAR-GARCH model outperforms the competitors for each variable and this is confirmed by the model confidence set. Moreover, both model combinations yield promising forecasts and are close to each other regarding overall forecasting accuracy (while beating the benchmark). These results are also confirmed by the Diebold-Mariano test, which highlights a significant difference in the performance of models with time-varying volatility compared to the homoskedastic benchmark.
Similar results are found also at the $20$th and $30$th percentiles.

Conversely, at the median ($\tau = 0.5$), the model confidence set shows no substantial evidence in favour of the inclusion of time-varying volatility; however, the DM test indicates that the QVAR-GARCH, as well as both model combinations, outperform the benchmark at the $10\%$ significance level for the Brent and CO$_2$ variables.
Regarding the right tail at the $70$th, $80$th, and $90$th percentiles, we find strong evidence in favour of time-varying volatility models and the model combination. In particular, the QVAR-GARCH model yields better performances for all the variables.

Table~\ref{tab:Result_rolling_horiz5} contains additional results for a longer forecasting horizon of $h=5$ days ahead. The main findings regarding the importance of accounting for time-varying volatility and the gains in accuracy against the QVAR benchmark are obtained. Instead, it is interesting to note that the best-performing model (and the one with larger weights in the model combination) in most scenarios is the QVAR-SV model.

As a robustness check, in the Supplement, we consider different time specifications for the out-of-sample analysis and for the time-varying weights but the results are in line with those provided in the paper.

Overall, these results highlight that accounting for time-varying volatility in its different specifications improves forecast accuracy, especially when investigating the tails. Moreover, the model combinations obtained using time-varying or average weights outperform the benchmark in every quantile/variable setting.
This suggests that when a forecaster is concerned with predicting the tail behaviour of a variable, accounting for time-varying volatility is crucial to obtain more reliable results.

%%%%%%%%%%%%%%%%%%%%%%%%%%%%%%%%%%%%%%%%%%%%%%%%%%%%%%%%%%%%%%%%%%%%%%%%%%%

\section{Conclusions}        \label{sec:conclusion}

Motivated by the increasing interest of central bankers, researchers, and policy institutions in tail risk evaluation and forecasting, we extended the multivariate quantile regression framework to account for alternative time-varying volatility specification through stochastic volatility or GARCH effects.
These models have been complemented by a model combination that exploits the quantile score to define the (possibly time-varying) weights.
To make inferences on the latent volatility paths, we designed an efficient MCMC algorithm to sample the entire path of the series-specific time-varying volatility jointly and independently across equations, allowing for meaningful computational savings.

We applied the proposed methods in multivariate formulations to forecast energy commodities at several quantile levels.
The main findings highlight that including time-varying volatility significantly improves the forecasting performance in all cases, especially on the tails of the distribution and at short and long horizons. Moreover, as no single model is found to be superior to others across quantiles or variables, we propose a multivariate model combination with a time-varying weighting scheme.

The proposed methods can be extended to more general time-varying parameter models by introducing a stochastic process over time for the quantile-specific coefficients, $\bbeta$. This would allow each covariate to affect the distribution of the desired indicator heterogeneously across quantiles and time and could improve the quantile forecasting performance.
Another interesting avenue of research concerns the design of more general observation-driven methods for the time-varying scale along the lines of score-driven models \citep{creal13generalized}.

%%%%%%%%%%%%%%%%%%%%%%%%%%%%%%%%%%%%%%%%%%
\bibliographystyle{chicago}
\bibliography{biblio}

\begin{thebibliography}{}

\bibitem[\protect\citeauthoryear{Aastveit, ter Ellen, and Mantoan}{Aastveit
  et~al.}{2024}]{aastveit2022quantile}
Aastveit, K.~A., S.~ter Ellen, and G.~Mantoan (2024).
\newblock {Quantile density combination: An application to US GDP forecasts}.
\newblock Technical Report~14, Norges Bank's Working Papers.

\bibitem[\protect\citeauthoryear{Adrian, Boyarchenko, and Giannone}{Adrian
  et~al.}{2019}]{adrian2019vulnerable}
Adrian, T., N.~Boyarchenko, and D.~Giannone (2019).
\newblock Vulnerable growth.
\newblock {\em American Economic Review\/}~{\em 109\/}(4), 1263--89.

\bibitem[\protect\citeauthoryear{Adrian, Grinberg, Liang, Malik, and Yu}{Adrian
  et~al.}{2022}]{adrian2020GaR}
Adrian, T., F.~Grinberg, N.~Liang, S.~Malik, and J.~Yu (2022).
\newblock The term structure of growth-at-risk.
\newblock {\em American Economic Journal: Macroeconomics\/}~{\em 14\/}(3),
  283--323.

\bibitem[\protect\citeauthoryear{Ando and Bai}{Ando and Bai}{2020}]{AB2020}
Ando, T. and J.~Bai (2020).
\newblock Quantile co-movement in financial markets: A panel quantile model
  with unobserved heterogeneity.
\newblock {\em Journal of the American Statistical Association\/}~{\em
  115\/}(529), 266--279.

\bibitem[\protect\citeauthoryear{Antolin-Diaz, Drechsel, and
  Petrella}{Antolin-Diaz et~al.}{2024}]{antolin2023advances}
Antolin-Diaz, J., T.~Drechsel, and I.~Petrella (2024).
\newblock {Advances in nowcasting economic activity: The role of heterogeneous
  dynamics and fat tails}.
\newblock {\em Journal of Econometrics\/}~{\em 238\/}(2), 105634.

\bibitem[\protect\citeauthoryear{Atchad{\'e} and Rosenthal}{Atchad{\'e} and
  Rosenthal}{2005}]{atchade2005adaptive}
Atchad{\'e}, Y.~F. and J.~S. Rosenthal (2005).
\newblock On adaptive {Markov} chain {Monte} {Carlo} algorithms.
\newblock {\em Bernoulli\/}~{\em 11\/}(5), 815--828.

\bibitem[\protect\citeauthoryear{Bollerslev}{Bollerslev}{1986}]{bollerslev1986GARCH}
Bollerslev, T. (1986).
\newblock Generalized autoregressive conditional heteroskedasticity.
\newblock {\em Journal of Econometrics\/}~{\em 31\/}(3), 307--327.

\bibitem[\protect\citeauthoryear{Carriero, Clark, and Marcellino}{Carriero
  et~al.}{2016}]{carriero2016SV}
Carriero, A., T.~E. Clark, and M.~Marcellino (2016).
\newblock {Common drifting volatility in large Bayesian VARs}.
\newblock {\em Journal of Business \& Economic Statistics\/}~{\em 34\/}(3),
  375--390.

\bibitem[\protect\citeauthoryear{Carriero, Clark, and Marcellino}{Carriero
  et~al.}{2022}]{carriero2022nowcasting}
Carriero, A., T.~E. Clark, and M.~Marcellino (2022).
\newblock Nowcasting tail risk to economic activity at a weekly frequency.
\newblock {\em Journal of Applied Econometrics\/}~{\em 37\/}(5), 843--866.

\bibitem[\protect\citeauthoryear{Chavleishvili and Manganelli}{Chavleishvili
  and Manganelli}{2024}]{Manganelli2021quantileIRF}
Chavleishvili, S. and S.~Manganelli (2024).
\newblock Forecasting and stress testing with quantile vector autoregression.
\newblock {\em Journal of Applied Econometrics\/}~{\em 39\/}(1), 66--85.

\bibitem[\protect\citeauthoryear{Chen, Dolado, and Gonzalo}{Chen
  et~al.}{2021}]{CDG2021}
Chen, L., J.~J. Dolado, and J.~Gonzalo (2021).
\newblock Quantile factor models.
\newblock {\em Econometrica\/}~{\em 89\/}(2), 875--910.

\bibitem[\protect\citeauthoryear{Clark, Huber, Koop, Marcellino, and
  Pfarrhofer}{Clark et~al.}{2024}]{clark2021investigating}
Clark, T.~E., F.~Huber, G.~Koop, M.~Marcellino, and M.~Pfarrhofer (2024).
\newblock {Investigating Growth-at-Risk Using a Multicountry Nonparametric
  Quantile Factor Model}.
\newblock {\em Journal of Business \& Economic Statistics\/}, 1--16.

\bibitem[\protect\citeauthoryear{Clark and Ravazzolo}{Clark and
  Ravazzolo}{2015}]{clark2015comparison}
Clark, T.~E. and F.~Ravazzolo (2015).
\newblock {Macroeconomic forecasting performance under alternative
  specifications of time-varying volatility}.
\newblock {\em Journal of Applied Econometrics\/}~{\em 30\/}(4), 551--575.

\bibitem[\protect\citeauthoryear{Cogley and Sargent}{Cogley and
  Sargent}{2005}]{cogley2005drifts}
Cogley, T. and T.~J. Sargent (2005).
\newblock {Drifts and volatilities: Monetary policies and outcomes in the post
  WWII US}.
\newblock {\em Review of Economic Dynamics\/}~{\em 8\/}(2), 262--302.

\bibitem[\protect\citeauthoryear{Cox, Gudmundsson, Lindgren, Bondesson,
  Harsaae, Laake, Juselius, and Lauritzen}{Cox
  et~al.}{1981}]{cox1981statistical}
Cox, D.~R., G.~Gudmundsson, G.~Lindgren, L.~Bondesson, E.~Harsaae, P.~Laake,
  K.~Juselius, and S.~L. Lauritzen (1981).
\newblock Statistical analysis of time series: Some recent developments.
\newblock {\em Scandinavian Journal of Statistics\/}, 93--115.

\bibitem[\protect\citeauthoryear{Creal, Koopman, and Lucas}{Creal
  et~al.}{2013}]{creal13generalized}
Creal, D., S.~J. Koopman, and A.~Lucas (2013).
\newblock Generalized autoregressive score models with applications.
\newblock {\em Journal of Applied Econometrics\/}~{\em 28\/}(5), 777--795.

\bibitem[\protect\citeauthoryear{Cross, Hou, Koop, and Poon}{Cross
  et~al.}{2023}]{cross2023large}
Cross, J.~L., C.~Hou, G.~Koop, and A.~Poon (2023).
\newblock Large stochastic volatility in mean vars.
\newblock {\em Journal of Econometrics\/}~{\em 236\/}(1), 105469.

\bibitem[\protect\citeauthoryear{D'Agostino, Gambetti, and Giannone}{D'Agostino
  et~al.}{2013}]{dAgostino2013SV}
D'Agostino, A., L.~Gambetti, and D.~Giannone (2013).
\newblock Macroeconomic forecasting and structural change.
\newblock {\em Journal of Applied Econometrics\/}~{\em 28\/}(1), 82--101.

\bibitem[\protect\citeauthoryear{Diebold and Mariano}{Diebold and
  Mariano}{1995}]{Diebold1995}
Diebold, F. and R.~Mariano (1995).
\newblock Comparing predictive accuracy.
\newblock {\em Journal of Business \& Economic Statistics\/}~{\em 13\/}(3),
  253--263.

\bibitem[\protect\citeauthoryear{Engle}{Engle}{1982}]{engle1982ARCH}
Engle, R.~F. (1982).
\newblock {Autoregressive conditional heteroscedasticity with estimates of the
  variance of United Kingdom inflation}.
\newblock {\em Econometrica\/}~{\em 50\/}(4), 987--1007.

\bibitem[\protect\citeauthoryear{Ferrara, Mogliani, and Sahuc}{Ferrara
  et~al.}{2022}]{ferrara2022Midas}
Ferrara, L., M.~Mogliani, and J.-G. Sahuc (2022).
\newblock High-frequency monitoring of growth at risk.
\newblock {\em International Journal of Forecasting\/}~{\em 38\/}(2), 582--595.

\bibitem[\protect\citeauthoryear{Giacomini and Komunjer}{Giacomini and
  Komunjer}{2005}]{giacomini2005evaluation}
Giacomini, R. and I.~Komunjer (2005).
\newblock Evaluation and combination of conditional quantile forecasts.
\newblock {\em Journal of Business \& Economic Statistics\/}~{\em 23\/}(4),
  416--431.

\bibitem[\protect\citeauthoryear{Gianfreda, Ravazzolo, and Rossini}{Gianfreda
  et~al.}{2023}]{gianfreda2023large}
Gianfreda, A., F.~Ravazzolo, and L.~Rossini (2023).
\newblock Large time-varying volatility models for hourly electricity prices.
\newblock {\em Oxford Bulletin of Economics and Statistics\/}~{\em 85\/}(3),
  545--573.

\bibitem[\protect\citeauthoryear{Hansen, Lunde, and Nason}{Hansen
  et~al.}{2011}]{hansen_etal.2011}
Hansen, P.~R., A.~Lunde, and J.~M. Nason (2011).
\newblock {The model confidence set}.
\newblock {\em Econometrica\/}~{\em 79}, 453--497.

\bibitem[\protect\citeauthoryear{Iacopini, Poon, Rossini, and Zhu}{Iacopini
  et~al.}{2023}]{iacopini2022bayesian}
Iacopini, M., A.~Poon, L.~Rossini, and D.~Zhu (2023).
\newblock {Bayesian mixed-frequency quantile vector autoregression: Eliciting
  tail risks of monthly US GDP}.
\newblock {\em Journal of Economic Dynamics and Control\/}~{\em 157}, 104757.

\bibitem[\protect\citeauthoryear{Kastner and Fr{\"u}hwirth-Schnatter}{Kastner
  and Fr{\"u}hwirth-Schnatter}{2014}]{kastner2014ancillarity}
Kastner, G. and S.~Fr{\"u}hwirth-Schnatter (2014).
\newblock {Ancillarity-sufficiency interweaving strategy (ASIS) for boosting
  MCMC estimation of stochastic volatility models}.
\newblock {\em Computational Statistics \& Data Analysis\/}~{\em 76}, 408--423.

\bibitem[\protect\citeauthoryear{Koenker and Bassett}{Koenker and
  Bassett}{1978}]{koenker1978regression}
Koenker, R. and G.~Bassett (1978).
\newblock Regression quantiles.
\newblock {\em Econometrica\/}, 33--50.

\bibitem[\protect\citeauthoryear{Korobilis and Schr{\"o}der}{Korobilis and
  Schr{\"o}der}{2022}]{korobilis2022probabilistic}
Korobilis, D. and M.~Schr{\"o}der (2022).
\newblock Probabilistic quantile factor analysis.
\newblock {\em arXiv preprint arXiv:2212.10301\/}.

\bibitem[\protect\citeauthoryear{Kotz, Kozubowski, and Podg{\'o}rski}{Kotz
  et~al.}{2001}]{kotz2001laplace}
Kotz, S., T.~Kozubowski, and K.~Podg{\'o}rski (2001).
\newblock {\em The Laplace distribution and generalizations: A revisit with
  applications to communications, economics, engineering, and finance}.
\newblock Number 183. Springer Science \& Business Media.

\bibitem[\protect\citeauthoryear{Kozumi and Kobayashi}{Kozumi and
  Kobayashi}{2011}]{kozumi2011gibbs}
Kozumi, H. and G.~Kobayashi (2011).
\newblock Gibbs sampling methods for {Bayesian} quantile regression.
\newblock {\em Journal of Statistical Computation and Simulation\/}~{\em
  81\/}(11), 1565--1578.

\bibitem[\protect\citeauthoryear{Marcellino, Porqueddu, and
  Venditti}{Marcellino et~al.}{2016}]{marcellino2016}
Marcellino, M., M.~Porqueddu, and F.~Venditti (2016).
\newblock Short-term {GDP} forecasting with a mixed-frequency dynamic factor
  model with stochastic volatility.
\newblock {\em Journal of Business \& Economic Statistics\/}~{\em 34\/}(1),
  118--127.

\bibitem[\protect\citeauthoryear{Petrella and Raponi}{Petrella and
  Raponi}{2019}]{petrella2019joint}
Petrella, L. and V.~Raponi (2019).
\newblock Joint estimation of conditional quantiles in multivariate linear
  regression models with an application to financial distress.
\newblock {\em Journal of Multivariate Analysis\/}~{\em 173}, 70--84.

\bibitem[\protect\citeauthoryear{Primiceri}{Primiceri}{2005}]{Primiceri2005}
Primiceri, G.~E. (2005).
\newblock {Time varying structural vector autoregressions and monetary policy}.
\newblock {\em The Review of Economic Studies\/}~{\em 72\/}(3), 821--852.

\bibitem[\protect\citeauthoryear{Ravazzolo and Rossini}{Ravazzolo and
  Rossini}{2023}]{Ravazzolo23}
Ravazzolo, F. and L.~Rossini (2023).
\newblock {Is the Price Cap for Gas Useful? Evidence from European Countries}.
\newblock Technical Report 023, FEEM Working Papers.

\bibitem[\protect\citeauthoryear{Taylor}{Taylor}{1986}]{taylor1986modelling}
Taylor, S.~J. (1986).
\newblock {\em Modelling financial time series}.
\newblock Wiley.

\end{thebibliography}

\end{document}